\documentclass[apj]{emulateapj}

\newcommand{\kms}{{km s$^{-1}$}~}
\newcommand{\kmsend}{{km s$^{-1}$}}
\newcommand{\rh}{{$r_\mathrm{h}$}~}
\newcommand{\vrad}{{$v_\mathrm{r}$}~}
\newcommand{\rgc}{{$R_\mathrm{GC}$}~}

\shorttitle{Spectroscopy of Virgo IGCs and UCDs}
\shortauthors{Ko et al.}

\begin{document}


\title{To the Edge of M87 and Beyond: Spectroscopy of Intracluster Globular Clusters and Ultra Compact Dwarfs in the Virgo Cluster}


\author{Youkyung Ko$^{1}$, Ho Seong Hwang$^{2}$, Myung Gyoon Lee$^{1}$, Hong Soo Park$^{3}$, \\Sungsoon Lim$^{5,6}$, Jubee Sohn$^{4}$, In Sung Jang$^{1,7}$, Narae Hwang$^{3}$, and Byeong-Gon Park$^{3}$}
\affil{$^{1}$Astronomy Program, Department of Physics and Astronomy, Seoul National University, 1 Gwanak-ro, Gwanak-gu, Seoul 08826, Korea\\
$^{2}$School of Physics, Korea Institute for Advanced Study, 85 Hoegiro, Dongdaemun-Gu, Seoul 02455, Korea\\
$^{3}$Korea Astronomy and Space Science Institute, 776 Daedeokdae-Ro, Yuseong-Gu, Daejeon 34055, Korea\\
$^{4}$Smithsonian Astrophysical Observatory, 60 Garden Street, Cambridge, 02138, USA\\
$^{5}$Department of Astronomy, Peking University, Beijing 100871, China\\
$^{6}$Kavli Institute for Astronomy and Astrophysics, Peking University, Beijing 100871, China \\
$^{7}$Leibniz-Institut  f{\"u}r Astrophysik Potsdam (AIP), An der
Sternwarte 16, D-14482, Potsdam, Germany}

\email{ykko@astro.snu.ac.kr, mglee@astro.snu.ac.kr}


\begin{abstract}
We present the results from a wide-field spectroscopic survey of globular clusters (GCs) in the Virgo Cluster.
We obtain spectra for 201 GCs and 55 ultracompact dwarfs (UCDs) using the Hectospec on the Multiple Mirror Telescope, and derive their radial velocities.
We identify 46 genuine intracluster GCs (IGCs), not associated with any Virgo galaxies, using the 3D GMM test on the spatial and radial velocity distribution.
They are located at the projected distance 200 kpc $\lesssim$ R $\lesssim$ 500 kpc from the center of M87.
The radial velocity distribution of these IGCs shows two peaks, one at \vrad = 1023 \kms associated with the Virgo main body, and another at \vrad = 36 \kms associated with the infalling structure.
The velocity dispersion of the IGCs in the Virgo main body is $\sigma_{\rm{GC}} \sim$ 314 \kmsend, 
which is smoothly connected to the velocity dispersion profile of M87 GCs, but much lower than that of dwarf galaxies in the same survey field, $\sigma_{\rm{dwarf}} \sim$ 608 \kmsend.
The UCDs are more centrally concentrated on massive galaxies, M87, M86, and M84.
The radial velocity dispersion of the UCD system is much smaller than that of dwarf galaxies.
Our results confirm the large-scale distribution of Virgo IGCs indicated by previous photometric surveys.
The color distribution of the confirmed IGCs shows a bimodality similar to that of M87 GCs. This indicates that most IGCs are stripped off from dwarf galaxies and some from massive galaxies in the Virgo.

\end{abstract}





\section{INTRODUCTION}

Central regions of galaxy clusters provide a dense environment where galaxies interact frequently.
This results in the abundant tidal debris in the intergalactic region.
The intracluster light of galaxy clusters is expected to be the tidal debris stripped from cluster galaxies, 
and it is commonly detected not even in nearby galaxy clusters but also z $\sim$ 1 galaxy clusters 
\citep{ada05, mih05, zib05, gzz05, kri06, kb07, bur12, mih15}.
It is also expected that the globular clusters (GCs) stripped from cluster galaxies populate the intergalactic region.
They are called intracluster globular clusters (IGCs) of which existence was proposed by various studies (White 1987; Muzzio 1987; West et al. 1995).
They do not belong to any individual galaxies, but wander between galaxies.
There have been several attempts to identify these IGCs in several galaxy clusters 
including Fornax \citep{bas03, ber07, sch08, sch10, dab16},
Coma \citep{pen11}, A1185 \citep{jor03, wes11}, A1689 \citep{ala13}, and A2744 \citep{lj16}.

The Virgo Cluster is an excellent target to investigate the IGCs because of its proximity.
Several studies reported the presence of IGCs from wide-field imaging surveys of GCs in the Virgo.
\citet{tam06} performed a wide-field photometric survey of GCs out to a galactocentric radius of $\sim$ 500 kpc from the most massive galaxy of the Virgo, M87.
They detected a marginal excess of the surface number density of GCs compared with model predictions, indicating the presence of IGCs.
\citet{lph10} presented a large-scale map of GCs covering the entire Virgo Cluster using the SDSS photometric data, 
and suggested that there are a significant number of IGCs in the Virgo (N $\sim$ 11900).
Recently, \citet{dur14} used the Next Generation Virgo Cluster Survey (NGVS; Ferrarese et al. 2012) data, showing the number density map of the Virgo GCs with finer scales than that of \citet{lph10}. 
Most of these studies found that the IGCs have relatively blue colors, consistent with the idea that they originate from low-mass dwarf galaxies.
This scenario is supported by HST/ACS photometry of the resolved stars in four IGC candidates in the Virgo \citep{wil07}; these IGC candidates contain metal-poor stars with [M/H] $<$ -1.

Spectroscopy is very efficient in removing foreground and background contamination in the photometric candidates of the IGCs, because radial velocities of the targets can be derived.
However, there is only one spectroscopic observation of IGCs in Virgo in the literature.
\citet{fdk08} obtained spectra of IGC candidates in the Virgo and confirmed only  three IGCs at $\sim$ 750--850 kpc from M87.
In the central region of the Virgo, it is also important to distinguish the IGCs from the  GCs bound to M87.

M87 is the most massive elliptical galaxy located in the central region of the Virgo, 
and it is known as a marginal cD galaxy of which cD halo begins to emerge at R =3$\arcmin- 7\arcmin$, corresponding to 15 -- 35 kpc \citep{liu05,kor09}.
\citet{mih05} found that the cD halo of M87 is stretched to about 200 kpc.
M87 hosts a significant number of GCs out to $\sim$ 200 kpc \citep{lph10}.
\citet{cot01}, \citet{str11}, and \citet{zhu14} performed spectroscopic surveys of GCs around M87, 
but their radial coverage is only at $R<150$ kpc which is not large enough to find IGCs.

The velocity dispersion of IGCs is expected to be larger than the GCs bound to individual galaxies 
because IGCs are governed by the gravitational potential of their host cluster.
\citet{str11} quantitatively discussed the presence of the IGCs with extreme radial velocities more than 1000 \kms relative to the system velocity of M87. These IGCs with extreme velocities can increase the velocity dispersion of IGCs.
Based on the number density of GCs beyond 40 kpc from M87 estimated by \citet{tam06}, \citet{str11} expected that there might exist 17-87 IGCs in their survey field.
Then they suggested that 16-67\% of the IGCs in their survey field may have extreme radial velocities, similar to the fractions of intracluster planetary nebulae (ICPNe) and Virgo galaxies with extreme velocities. However, \citet{str11} could not identify individual IGCs, because their survey field covered only the M87 region at $R<150$ kpc.
On the other hand, \citet{cal14} reported a hypervelocity globular cluster with $v_r  \sim -1025$ \kmsend. However, they concluded that it is not a Virgo IGC, but an object ejected from the Virgo.

In this paper, we present a wide-field ($2^\circ \times 2^\circ$) spectroscopic survey of Virgo IGCs and UCDs with MMT/Hectospec.
This paper is organized as follows. 
We briefly describe the spectroscopic observation and data reduction in \S2.
In \S3, we select the genuine GCs in the Virgo, distinguishing them from the contaminants such as foreground stars and background galaxies.
We identify IGCs using spatial and kinematic information, and compare their kinematics with other tracers including ultra-compact dwarfs (UCDs), dwarf galaxies, and M87 GCs.
We discuss and summarize the results in \S4 and \S5, respectively.
We adopted a distance to M87 of 16.5 Mpc \citep{bla09}.
One arcmin corresponds to 4.80 kpc at the distance to M87. 
The heliocentric radial velocity of M87 is 1260 \kms \citep{kim14}.

\section{OBSERVATION AND DATA REDUCTION}
 	\subsection{Target Selection for Spectroscopy}

\begin{deluxetable*}{c c c c c c}
\tablecaption{Observation Log for the MMT/Hectospec Run \label{tab:obs.field}}
\tablewidth{\textwidth}
\tablehead{
\colhead{Mask Name} & \colhead{$\alpha$ (J2000)} & \colhead{$\delta$ (J2000)} & \colhead{N$^{a}$}
& \colhead{Exp. time} & \colhead{Date(UT)}
}
\startdata
Virgo-F1 & 12:25:13.66 & +12:59:43.8 & 258 & 5 $\times$ 1440 s & Feb 25, 2014 \\
Virgo-F2 & 12:25:25.00 & +11:58:55.3 & 255 & 5 $\times$ 1440 s & Feb 26, 2014 \\
Virgo-F3 & 12:25:28.50 & +12:28:38.1 & 255 & 5 $\times$ 1440 s & Feb 27, 2014 \\
Virgo-F4 & 12:28:32:69 & +12:02:38.3 & 258 & 5 $\times$ 1440 s & Mar 24, 2014 \\
Virgo-F5 & 12:28:48.44 & +12:52:53.6 & 258 & 5 $\times$ 1440 s & Mar 25, 2014 
\enddata
\tablecomments{$^{a}$ Number of object fibers among 300 fibers in each field. The remaining fibers are assigned to sky regions.}
\end{deluxetable*}

We selected GC candidates using the NGVS archival images covering the central region of the Virgo Cluster.
The NGVS is a wide-field imaging survey of the Virgo Cluster using MegaCam with a field of view of $1^{\circ} \times 1^{\circ}$ attached at the Canada-French-Hawaii Telescope \citep{fer12}.
	\begin{figure}[t]
\epsscale{1}
\includegraphics[trim=50 0 30 30,clip,width=\columnwidth]{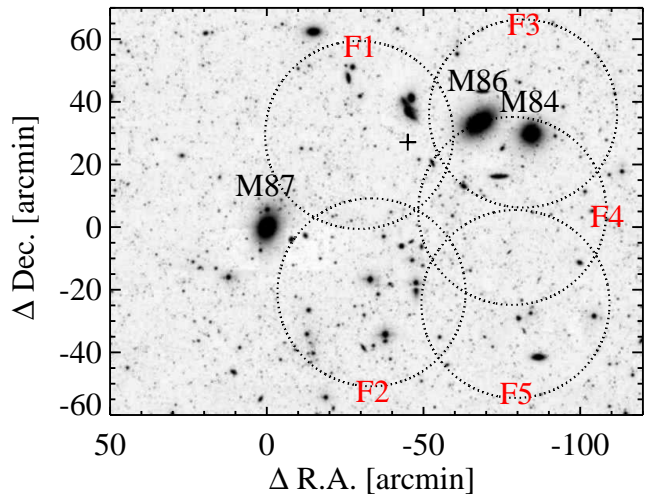}
\caption{Field configuration for the MMT/Hectospec observation overlaid on the gray-scale map of the SDSS.
A dotted-line circle represents a field-of-view of MMT/Hectospec with 1 deg diameter. The cross mark indicates the center of the Virgo Cluster derived from the number density map of dwarf galaxies in the Virgo \citep{bst85}. \label{fig:field_conf}}
	\end{figure}
The calibrated images were retrieved, covering about $3^{\circ} \times 2^{\circ}$ field, where M86, M84, and a part of M87 are included, which is shown in {\bf Figure \ref{fig:field_conf}}.
We did astrometry using SCAMP \citep{ber06} for these images and combined them using SWarp \citep{ber02} to make mosaic images in each filter.

We used SExtractor \citep{ba96} to detect the sources with a threshold of 3$\sigma$ and performed aperture photometry with an aperture radius of 1$\arcsec$.
Because most of GCs at the distance of the Virgo Cluster appear as point sources in the NGVS images, we selected point sources as GC candidates using the ``CLASS$\_$STAR" (stellarity index) parameter of SExtractor.
The stellarity index ranges from 0 to 1, which comes from the comparison between FWHM of each source and a given seeing size.
If this value of a given source is close to 1, the source is thought to be a point source.
We used a generous criterion to select many spectroscopic targets: i.e. stellarity index $\geqq$ 0.35 in $gri$ filters.
For these point sources, we did standard calibration using the SDSS DR12 $gri$ PSF photometry \citep{ala15} after transforming the SDSS to the CFHT/Megacam filter system\footnote{http://www.cadc-ccda.hia-iha.nrc-cnrc.gc.ca/en/megapipe/\\docs/filt.html}.
%
We used foreground galactic extinction values for the Virgo derived by \citet{sf11}.  
Hereafter, all photometry results are based on the AB magnitude and CFHT/MegaCam filter system.

	\begin{figure}[ht]
\epsscale{1.4}
\includegraphics[trim=70 0 90 0,clip,width=\columnwidth]{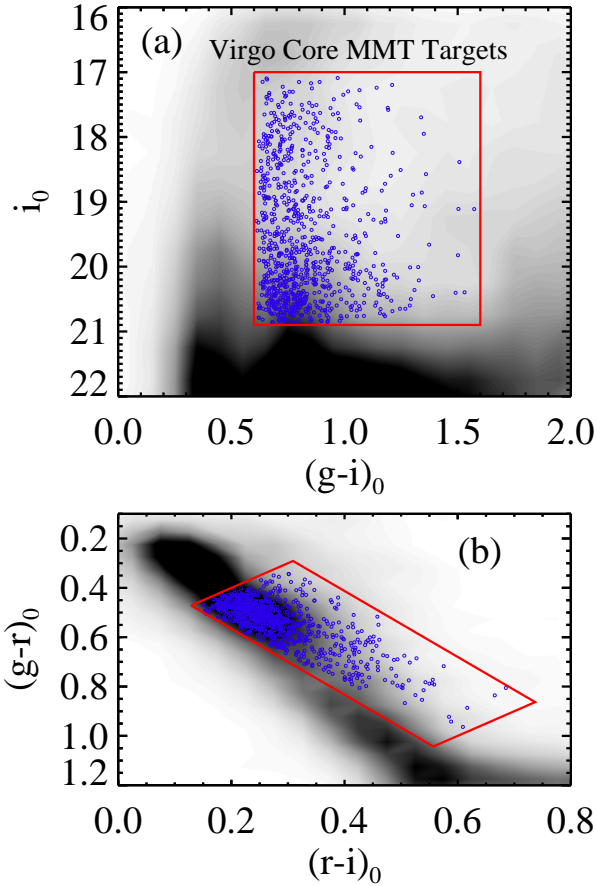}
\caption{Target selection for spectroscopy. 
(a) $i_0 -(g-i)_0 $ CMD and (b) $(g-r)_0 -(r-i)_0$ CCD of the point sources in $3^{\circ} \times 2^{\circ}$ field around the center of the Virgo derived from the NGVS data. The blue circles inside the red-line boxes represent spectroscopic targets observed in this study. Gray scales represent the Hess diagram.
\label{fig:cmd_ccd}}
	\end{figure}
\textbf{Figure \ref{fig:cmd_ccd}} shows the $i_0 -(g-i)_0$ color-magnitude diagram (CMD) and $(g-r)_0 -(r-i)_0$ color-color diagram (CCD) for the point sources in the field of $3^{\circ} \times 2^{\circ}$ around the central region of the Virgo.
Typical GCs in Virgo have a color range of $0.6 < g-i < 1.25$ \citep{dur14}.
We used a color range, $0.6 < (g-i)_0 < 1.6$ for target GC selection. We adopted a redder color limit ($(g-i)_0=1.6$) to include more spectroscopic GC targets for filling the available fibers.
We also used the $(g-r)_0 - (r-i)_0$ color-color diagram to select the GC candidates following the color criteria used for selecting M31 GCs in \citet{pea10}. 
\citet{pea10} provided SDSS photometry of the M31 GCs in the AB magnitude system so that we transformed it to the system in this paper.
The selection color criteria are as follows:
\begin{displaymath}
(g-r)_0 < 1.33 \times ((r-i)_0 - 0.39) + 0.82\rm{,~and}
\end{displaymath}
\begin{displaymath}
(g-r)_0 > 1.33 \times ((r-i)_0 - 0.63) + 0.72.
\end{displaymath}
We also set the magnitude limit ($17.0 < i_0 < 20.9$) to remove possible contamination by bright foreground stars and to avoid too faint targets.
We then visually inspected all the targets to remove background galaxies.
We included 11 additional GC candidates that have resolved HST/ACS photometric data \citep{lee11}.
In total, we selected 910 GC candidates for spectroscopic observations.

	\subsection{Spectroscopic Observation}

We carried out spectroscopic observation of GC candidates in the Virgo using the Hectospec \citep{fab05} mounted on the 6.5m Multiple-Mirror telescope in queue mode under the program ID 2014A-UAO-G18 (PI: Myung Gyoon Lee) between 2014 February and March.
The Hectospec is a 300 fiber fed optical spectrograph with the circular field of view of 1 degree in diameter.
We selected a 270 mm$^{-1}$ grating with a dispersion of 1.2 \rm{\AA} pixel$^{-1}$, covering the wavelength range of 3650 \rm{\AA} to 9200 \rm{\AA}.
We made five different configurations, covering the central region of the Virgo, as shown in \textbf{Figure \ref{fig:field_conf}}. The field coordinates and exposure times are given in \textbf{Table \ref{tab:obs.field}}.
We used 5 $\times$ 24 minute exposures for each field.

    \subsection{Data Reduction and Radial Velocity Measurement}

We used the version 2.0 of the HSRED reduction pipeline\footnote{This is an updated reduction pipeline originally developed by Richard Cool, and more details are at http://www.mmto.org/node/536.} for data reduction.
The pipeline includes bias and dark correction, flat-fielding, aperture extraction of spectra, and wavelength calibration.
The flux calibration was done following the steps in \citet{fab08} that include atmospheric extinction correction, Hectospec relative throughput correction, and absolute flux normalization with SDSS $r$ photometry.
The spectral resolution is $\sim$ 6 {\rm{\AA}}.
The median signal-to-noise ratios of the spectra of GC candidates at 3700 {\rm{\AA}} -- 7000 {\rm{\AA}} range from 4 to 72.
We derived heliocentric radial velocities of GC candidates using xcsao task in IRAF RVSAO package \citep{km98}.
We used 10 templates including the spectra of an A star, three M31 GCs, an SDSS QSO, elliptical, and spiral galaxies in the RVSAO package.
To apply the cross-correlation method \citep{td79}, we used absorption features of target spectra and templates over the wavelength range of 3800 -- 5400 {\rm{\AA}} where there are no strong sky emission lines, but are prominent absorption lines.
We adopted the radial velocities of targets with the highest correlation coefficient, 
and visually inspected all spectra with the absorption lines shifted as their derived radial velocities.
We matched the targets with \vrad $>$ 3000 \kms with galaxy spectra and the others with GC spectra.
The radial velocity uncertainties are derived by the correlation error, and their mean value is 28 \kmsend.
For 19 out of 910 targets, we could not determine their radial velocities because of their low signal-to-noise ratio ($<5$).
The radial velocities of targets derived in this study agree well with those from the literature (see \S3.1.3).\\

\section{RESULTS}
    \subsection{Construction of GC and UCD samples}
    	\subsubsection{Identification of GCs and UCDs from the spectroscopic sample}
	
We distinguish GCs from other contaminants using several criteria among spectroscopic targets.
The possible contaminants and their classification criteria are 1) background galaxies with \vrad $>$ 3000 \kmsend,
2) UCDs with \vrad $<$ 3000 \kms and \rh $>$ 9.5 pc, and
3) foreground stars with \vrad $<$ 500 \kms and their marked spectral features.
The detailed classification procedure of each step is described as follows.
  
First, we classify the targets with \vrad $>$ 3000 \kms as background galaxies. \citet{kim14} presented the Extended Virgo Cluster catalog (EVCC), and showed that there is a dip at \vrad $\sim$ 3000 \kms in the velocity distribution of the galaxies in the EVCC, separating the galaxies in the Virgo Cluster from background galaxies.
Among spectroscopic targets, there are 93 background galaxies. They have the radial velocities of 23427 km s$^{-1} <$ \vrad $<$ 240257 \kmsend.

	\begin{figure}[b]
\epsscale{1}
\includegraphics[trim=50 15 20 80,clip,width=\columnwidth]{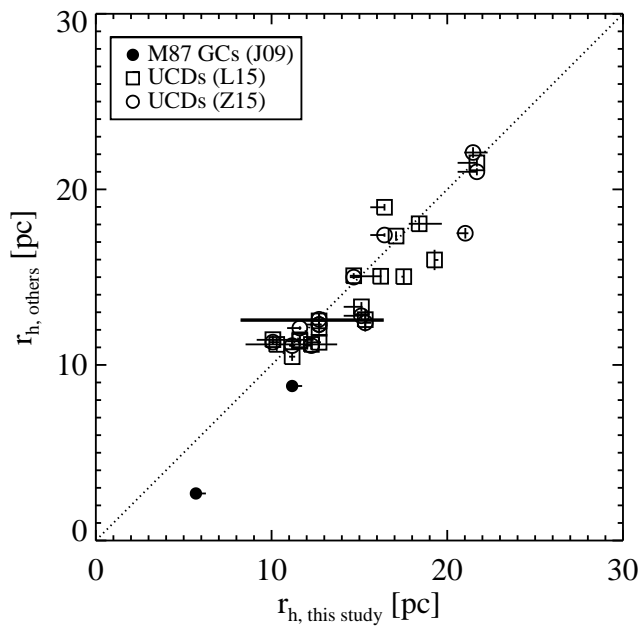}
\caption{Comparison of the half-light radii derived in this study with those derived in previous studies. 
The filled circles, open circles, and open squares represent the GCs in M87 \citep{jor09}, the UCDs in M87 \citep{zha15}, and the UCDs in M87 \citep{liu15}, respectively.
The dotted line denotes the one-to-one relation. \label{fig:size_comp}}
	\end{figure}

Second, we use the NGVS $i$ band images to measure the sizes of 798 GC candidates at \vrad $<$ 3000 \kms with the ISHAPE \citep{lar99} to distinguish GCs from UCDs.
The PSF FWHMs of the NGVS images we used are in the range 2.3-4.8 pixels, corresponding to 0$\rlap{.}{\arcsec}$4-0$\rlap{.}{\arcsec}$9. 
The ISHAPE manual suggests that the size estimates are reliable only when the FWHMs are larger than $\sim$ 0.1 $\times$ FWHM$_{\rm{PSF}}$.
We estimate the FWHM of 106 GC candidates adopting the KING30 function.
The half-light radii ($r_\mathrm{h}$) are derived by multiplying the FWHM by 1.48 following the ISHAPE manual.
They range from 5.3 pc to 33.7 pc, assuming that the distance to the Virgo is 16.5 Mpc.
These are larger than the mean half-light radius of the GCs in the Virgo galaxies (2.9 $\pm$ 1.2 pc) derived from HST/ACS images \citep{jor09}.
It is because we cannot derive the half-light radii of the GCs smaller than 4.9 pc from the ground-based images that have the limited resolution.
\textbf{Figure \ref{fig:size_comp}} shows a comparison of the half-light radii of GCs and UCDs derived in this study with those from the literature \citep{jor09, liu15, zha15}.
The size measurements of two GCs show a difference between this study and \citet{jor09}. We used ISHAPE on the NGVS images for the size measurement, while \citet{jor09} used KINGPHOT on the HST/ACS images. Thus this difference for the GCs is mainly due to the difference in the image resolution and the size measuring softwares.
However, for most of the UCDs, their sizes are well consistent with those from \citet{liu15} and \citet{zha15} who used KINGPHOT on the NGVS images. Some outliers of the UCDs are due to the difference in the size measuring softwares between this study and other studies.
Following the criterion in \citet{str11}, we classify the targets with \rh $>$ 9.5 pc as UCDs.
We find 55 UCDs from our sample.

\begin{figure*}
\epsscale{0.85}
\plotone{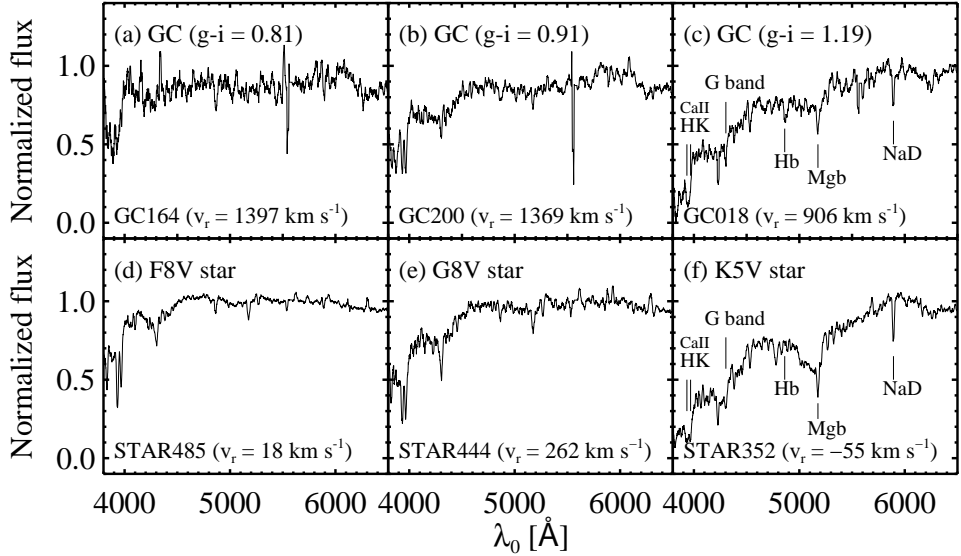}
\caption{Sample spectra of GCs and foreground stars.
(a) a GC (ID: GC164) with $i = 20.459$ mag and $(g-i) = 0.81$; 
(b) a GC (ID: GC200) with $i = 19.371$ mag and $(g-i) = 0.91$; 
(c) a GC (ID: GC018) with $i = 20.471$ mag and $(g-i) = 1.19$;
and (d), (e), and (f) stars with $i$ = 19.481, 20.041 and 18.427 mag classified as F8V (ID: STAR485), G8V (ID: STAR444), and K5V (ID: STAR352) stars by the HSRED package, respectively. 
All flux-calibrated spectra are plotted in the rest frame, smoothed using a boxcar filter with a size of 12 {\rm{\AA}}, and normalized at 5800 -- 5850 {\rm{\AA}}. \label{fig:sam_spectra}}
	\end{figure*}

Third, we remove foreground stars using size estimates, radial velocities, and spectral features.
Among 743 GC candidates, we classify 51 objects that have half-light radii of 5.3 pc $<$ \rh $<$ 9.5 pc as GCs in the Virgo because they are clearly extended than the PSF.
Among the rest of GC candidates, we select 95 GC candidates with 500 km s$^{-1} <$ \vrad $<$ 3000 \kms as genuine GCs in the Virgo.
Galactic stars rarely have the radial velocities greater than 500 \kmsend.
In addition, three of the co-authors independently conduct visual inspection of the spectra of the remaining 597 objects with \vrad $<$ 500 \kmsend.
We compare the spectral features in the target spectra with those in template spectra.
We use the spectra of the GCs with 500 km s$^{-1} <$ \vrad $<$ 3000 \kms in this study and those of Galactic stars in \citet{san01} as templates.
\textbf{Figure \ref{fig:sam_spectra}} shows the spectra of GCs and Galactic stars classified in this study.
For GC spectra, the continuum slope between 4000 \rm{\AA} and 7000 \rm{\AA} differs depending on the GC color.
Compared with GC spectra, the Mgb absorption line is generally broader and stronger in the spectra of stars, especially for dwarf stars.
The visual inspection results in a sample of 55 GCs with \vrad $<$ 500 \kmsend.
Finally, we obtain 201 GCs, 55 UCDs, 542 foreground stars, and 93 galaxies from 910 spectroscopic targets. 
\textbf{Table \ref{tab:star_gal.prop}} lists foreground stars and background galaxies observed in this study, 
and \textbf{Tables \ref{tab:gc.prop} and \ref{tab:ucd.prop}} list photometric and spectroscopic parameters of the GCs and UCDs, respectively.

\begin{deluxetable*}{l l l c c c c}
\tablecaption{Spectroscopic and photometric properties of foreground stars and background galaxies \label{tab:star_gal.prop}}
\tablewidth{\textwidth}
\tablehead{
\colhead{ID} & \colhead{$\alpha$ (J2000)} & \colhead{$\delta$ (J2000)} & \colhead{$i$} & \colhead{$g-r$} & \colhead{$g-i$} & \colhead{$v_\mathrm{r}$}\\
\colhead{} & \colhead{[deg]} & \colhead{[deg]} & \colhead{[mag]} & \colhead{[mag]} & \colhead{[mag]} & \colhead{[\kmsend]}
}
\startdata
STAR001 & 185.848618 & 12.998183 & 19.515 $\pm$ 0.002 & 0.456 $\pm$ 0.002 & 0.685 $\pm$ 0.002 & --33 $\pm$ 15 \\
STAR002 & 185.851730 & 13.125706 & 18.030 $\pm$ 0.001 & 0.533 $\pm$ 0.001 &  0.759 $\pm$ 0.001 & 18 $\pm$ 5 \\
STAR003 & 185.853333 & 13.100986 & 19.770 $\pm$ 0.002 & 0.465 $\pm$ 0.003 & 0.703 $\pm$ 0.002 & --33 $\pm$ 20 \\
STAR004 & 185.857193 & 12.048497 & 17.782 $\pm$ 0.001 & 0.502 $\pm$ 0.001 & 0.776 $\pm$ 0.001 & --95 $\pm$ 9 \\
STAR005 & 185.866241 & 12.020297 & 20.861 $\pm$ 0.006 & 0.551 $\pm$ 0.006 & 0.835 $\pm$ 0.008 & 68 $\pm$ 25 \\
\hline
GAL01 & 185.852463 & 12.890166 & 20.032 $\pm$ 0.002 & 0.724 $\pm$ 0.004 & 1.167 $\pm$ 0.003 & 58978 $\pm$ 7 \\
GAL02 & 185.852768 & 12.977111 & 19.596 $\pm$ 0.002 & 0.621 $\pm$ 0.003 & 1.087 $\pm$ 0.002 & 61396 $\pm$ 10 \\
GAL03 & 185.866852 & 12.093072 & 20.669 $\pm$ 0.005 & 0.605 $\pm$ 0.005 & 0.920 $\pm$ 0.007 & 240245 $\pm$ 186 \\
GAL04 & 185.875107 & 12.850986 & 20.750 $\pm$ 0.004 & 0.575 $\pm$ 0.005 & 0.827 $\pm$ 0.005 & 76235 $\pm$ 4 \\
GAL05 & 185.988205 & 13.227654 & 20.826 $\pm$ 0.004 & 0.537 $\pm$ 0.006 & 0.886 $\pm$ 0.006 & 69472 $\pm$ 4
\enddata
\tablecomments{Table \ref{tab:star_gal.prop} is published in its entirety in the electronic edition.  
The five sample foreground stars and background galaxies are shown here regarding its form and content.}

\end{deluxetable*}

\begin{turnpage}
\begin{deluxetable*}{l l l c c c c c c c c c c}
\tablecaption{Spectroscopic and photometric properties of the combined GC sample \label{tab:gc.prop}}
\tablewidth{0pt}
\tablehead{
\colhead{ID} & \colhead{$\alpha$ (J2000)} & \colhead{$\delta$ (J2000)} & \colhead{$i$} & \colhead{$g-r$} & \colhead{$g-i$} & \colhead{Phot.} & \colhead{$v_\mathrm{r}$} & \colhead{$v_\mathrm{r}$} & \colhead{$r_\mathrm{h}$} & \colhead{$r_\mathrm{h}$} & \colhead{Host} & \colhead{Host}\\
\colhead{} & \colhead{[deg]} & \colhead{[deg]} & \colhead{[mag]} & \colhead{[mag]} & \colhead{[mag]} & \colhead{Src.$^a$} & \colhead{[\kmsend]} & \colhead{Src.} & \colhead{[pc]} & \colhead{Src.} & \colhead{{\sc Mclust$^b$}} & \colhead{Rv cut}
}
\startdata
GC001 & 185.872772 & 13.161441 & 19.751 $\pm$ 0.002 & 0.557 $\pm$ 0.003 & 0.869 $\pm$ 0.002 & K16 & 1137 $\pm$ 22 & K16 & 8.3$^{+1.0}_{-0.2}$ & K16 & IGC (H) & IGC \\
GC002 & 185.945038 & 12.503950 & 20.254 $\pm$ 0.003 & 0.488 $\pm$ 0.004 & 0.727 $\pm$ 0.003 & K16 & 840 $\pm$ 51 & K16 & 6.8 $^{+0.1}_{-0.2}$ & K16 & IGC (H) & IGC \\
GC003 & 185.950409 & 13.076539 & 20.675 $\pm$ 0.004 & 0.546 $\pm$ 0.005 & 0.820 $\pm$ 0.005 & K16 & 124 $\pm$ 26 & K16 & - & - & IGC (H) & IGC \\
GC004 & 185.970627 & 12.773033 & 19.651 $\pm$ 0.002 & 0.552 $\pm$ 0.003 & 0.828 $\pm$ 0.002 & K16 & 1167 $\pm$ 28 & K16 & 7.0 $^{+0.3}_{-0.2}$ & K16 & M84 (F) & IGC \\
GC005 & 185.975204 & 12.940945 & 20.595 $\pm$ 0.004 & 0.532 $\pm$ 0.005 & 0.829 $\pm$ 0.004 & K16 & 796 $\pm$ 47 & K16 & 6.6 $^{+1.2}_{-<0.1}$ & K16 & M84 (F) & IGC \\
\hline
H52724 & 187.299957 & 12.466810 & 20.75 & 0.52 & 0.83 & S11 & 1124 $\pm$ 15 & S11 & - & - & M87 (C) & - \\
H46835 & 187.313446 & 12.411310 & 21.32 & 0.62 & 0.80 & S11 & 1253 $\pm$ 17 & S11 & - & - & M87 (C) & - \\
H57011 & 187.331863 & 12.510560 & 20.54 & 0.68 & 0.90 & S11 & 1409 $\pm$ 11 & S11 & - & - & M87 (C) & - \\
H44278 & 187.377747 & 12.390920 & 21.14 & 0.54 & 0.83 & S11 & 1511 $\pm$ 27 & S11 & - & - & M87 (C) & - \\
H58443 & 187.411087 & 12.526390 & 21.46 & 0.67 & 0.73 & S11 & 441 $\pm$ 24 & S11 & - & - & IGC (H) & - 
\enddata
\tablecomments{Table \ref{tab:gc.prop} is published in its entirety in the electronic edition.  
The five sample GCs confirmed in this study and five sample GCs from \citet{str11} are shown here regarding its form and content.}
\tablenotetext{a}{The magnitudes of this study and \citet{str11} are CFHT/MegaCam AB and dereddened SDSS AB magnitudes, respectively.}
\tablenotetext{b}{The {\sc Mclust} subgroup that a given GC belong to (see Table \ref{tab:gc_ucd_sub}) is in the parenthesis.}
\end{deluxetable*}
\begin{deluxetable*}{l l l c c c c c c c c c}
\tablecaption{Spectroscopic and photometric properties of the combined UCD sample \label{tab:ucd.prop}}
\tablewidth{0pt}
\tablehead{
\colhead{ID} & \colhead{$\alpha$ (J2000)} & \colhead{$\delta$ (J2000)} & \colhead{$i$} & \colhead{$g-r$} & \colhead{$g-i$} & \colhead{Phot.} & \colhead{$v_\mathrm{r}$} & \colhead{$v_\mathrm{r}$} & \colhead{$r_\mathrm{h}$} & \colhead{$r_\mathrm{h}$} & \colhead{Mclust} \\
\colhead{} & \colhead{[deg]} & \colhead{[deg]} & \colhead{[mag]} & \colhead{[mag]} & \colhead{[mag]} & \colhead{Src.$^a$} & \colhead{[\kmsend]} & \colhead{Src.} & \colhead{[pc]} & \colhead{Src.} & \colhead{Group$^b$}
}
\startdata
UCD01 & 185.919037 & 12.211850 & 20.379 $\pm$ 0.004 & 0.680 $\pm$ 0.005 & 1.143 $\pm$ 0.006 & K16 & 867 $\pm$ 17 & K16 & 11.4$^{+0.1}_{-0.1}$ & K16 & C \\
UCD02 & 186.069351 & 13.008061 & 20.727 $\pm$ 0.004 & 0.503 $\pm$ 0.005 & 0.819 $\pm$ 0.005 & K16 & 719 $\pm$ 39 & K16 & 9.9$^{+0.6}_{-<0.1}$ & K16 & C \\
UCD03 & 186.190231 & 13.049951 & 19.961 $\pm$ 0.002 & 0.622 $\pm$ 0.003 & 1.005 $\pm$ 0.003 & K16 & 952 $\pm$ 22 & K16 & 20.8$^{+1.6}_{-0.1}$ & K16 & C \\
UCD04 & 186.192642 & 12.341794 & 19.766 $\pm$ 0.003 & 0.503 $\pm$ 0.003 & 0.808 $\pm$ 0.003 & K16 & 955 $\pm$ 19 & K16 & 10.7$^{+0.1}_{-0.4}$ & K16 & C\\
UCD05 & 186.242004 & 12.895044 & 20.096 $\pm$ 0.002 & 0.621 $\pm$ 0.004 & 0.983 $\pm$ 0.003 & K16 & 1219 $\pm$ 23 & K16 & 18.4$^{+0.8}_{-0.9}$ & K16 & C \\
\hline
M87UCD-29 & 187.027588 & 12.410120 & 20.21 & 0.52 & 0.73 & Z15 & 599 $\pm$ 33 & Z15 & 12.6 $\pm$ 0.3 & Z15 & A \\
M87UCD-34 & 187.311966 & 11.895510 & 19.40 & 0.54 & 0.79 & Z15 & 905 $\pm$ 19 & Z15 & 11.7 $\pm$ 0.3 & Z15 & A \\	
M87UCD-20 & 187.422165 & 12.664570 & 19.45 & 0.52 & 0.80 & Z15 & 1754 $\pm$ 105 & Z15 & 16.7 $\pm$ 0.3 & Z15 & A \\
M87UCD-22 & 187.467834 & 12.627160 & 19.92 & 0.59 & 0.88 & Z15 & 905 $\pm$ 20 & Z15 & 11.7$\pm$ 0.2 & Z15 & A \\
M87UCD-10 & 187.508118 & 12.707470 & 19.17 & 0.45 & 0.69 & Z15 & 1178 $\pm$ 30 & Z15 & 21.9 $\pm$ 0.3 & Z15 & A 
\enddata
\tablecomments{Table \ref{tab:ucd.prop} is published in its entirety in the electronic edition.  
The five sample UCDs confirmed in this study and five sample UCDs from \citet{zha15} are shown here regarding its form and content.}
\tablenotetext{a}{The magnitudes of this study and \citet{str11} are CFHT/MegaCam AB and dereddened SDSS AB magnitudes, respectively.}
\tablenotetext{b}{The {\sc Mclust} subgroup that a given UCD belong to (see Table \ref{tab:gc_ucd_sub}) is in the parenthesis.}
\end{deluxetable*}
\end{turnpage}

\begin{figure}
\epsscale{1}
\includegraphics[trim=60 20 60 25,clip,width=\columnwidth]{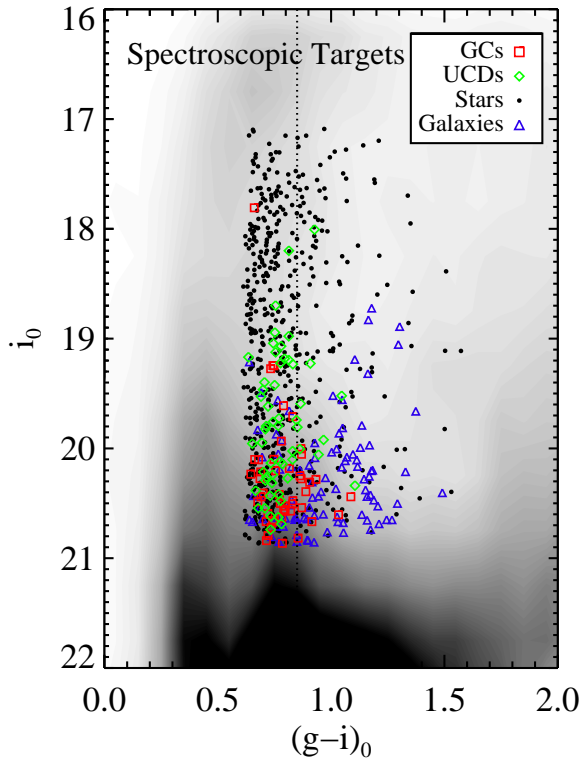}
\caption{$i_0 - (g-i)_0$ CMD of the spectroscopic targets classified in this study. Gray scales are the same as Figure \ref{fig:cmd_ccd}(a).
The squares, diamonds, circles, and triangles represent GCs, UCDs, foreground stars, and galaxies, respectively.
The dotted line is $(g-i)_0 = 0.8$ which is a criterion to divide GCs into blue and red GCs.
\label{fig:cmd}}
	\end{figure}

\textbf{Figure \ref{fig:cmd}} shows the $i_0 - (g-i)_0$ CMD of GCs, UCDs, foreground stars, and galaxies in the spectroscopic targets.
The majority of the GCs and UCDs have a color range of $0.6 < (g-i)_0 < 1.0$.
Most of red objects with $(g-i)_0 > 1.0$ turn out to be foreground stars or background galaxies.
We will discuss in detail the color distribution of GCs in \S4.2.
The $i_0$ magnitudes of most of GCs, UCDs and background galaxies are in the range 18.7 $< i_0 <$ 20.8, but those of stars are in the range 17.0 $ < i_0 < $ 20.8.
One GC with $i_0$ = 17.8 (ID: GC068) is as much as 1.5 mag brighter than other GCs, while it has the half-light radius similar to other GCs confirmed in this study (\rh $\sim$ 6 pc). 
Consequently, its luminosity density is $\sim 5 \times 10^{4} L_{i,\odot}$ pc$^{-2}$, which is higher than those of other GCs and UCDs ($\sim$ a few thousands $L_{i,\odot}$ pc$^{-2}$). 
\citet{mis11} and \citet{san15} reported the existence of some dense stellar systems in the UCD and compact cluster samples. 
The luminosity densities of HUCD1 in Hydra I \citep{mis11}, M59-UCD3, and M85-HCC1 \citep{san15} are $\sim 4.8 \times 10^{3} L_{V,\odot}$ pc$^{-2}$, $\sim 2.7 \times 10^{4} L_{r,\odot}$ pc$^{-2}$, and $\sim 4 \times 10^{5} L_{r,\odot}$ pc$^{-2}$, respectively. 
The luminosity density of GC068 in this study is comparable with that of M59-UCD3, five times higher than HUCD1, and one tenth of that of M85-HCC1.
	
\begin{figure}
\includegraphics[trim=32 27 40 15,clip,width=\columnwidth]{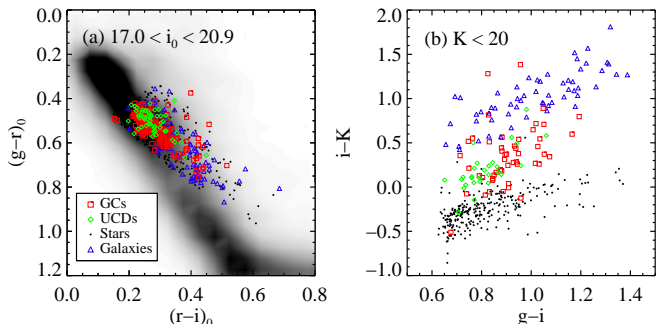}
\caption{(a) $(g-r)_0-(r-i)_0$ CCD of the spectroscopic targets classified in this study. Gray scales represent the same as Figure \ref{fig:cmd_ccd}(b).
(b) $(i-K)-(g-i)$ CCD of the spectroscopic targets with $K < 20$. The petrosian magnitudes are used in this plot.
The symbols are the same as Figure \ref{fig:cmd}.
\label{fig:gc_ccd}}
	\end{figure}

\textbf{Figure \ref{fig:gc_ccd}} shows the $(g-r)_0 -(r-i)_0$ and $(i-K)-(g-i)$ CCDs for the spectroscopic targets.
Many GCs and UCDs are in the color ranges of $0.4 < (g-r)_0 < 0.65$ and $0.2 < (r-i)_0 < 0.35$.
The colors of different populations are significantly distinct in the optical to near-infrared CCD \citep{mun14}.
We derive photometry for the point sources in the $K$ band images taken from the UKIRT/WFCAM, and use the Petrosian AB magnitudes to construct the CCD.
The GCs and UCDs occupy a similar color domain, but the foreground stars and galaxies are clearly separated from the GCs and UCDs in the $(i-K) - (g-i)$ CCD.
There are two GCs in the stellar sequence (ID: GC041 and GC068).
They are classified as GCs because GC068 is an extended source (\rh = 6 pc), and GC041 has the radial velocity of \vrad = 812 \kmsend.

\begin{figure}
\epsscale{1}
\includegraphics[trim=0 20 0 5,clip,width=\columnwidth]{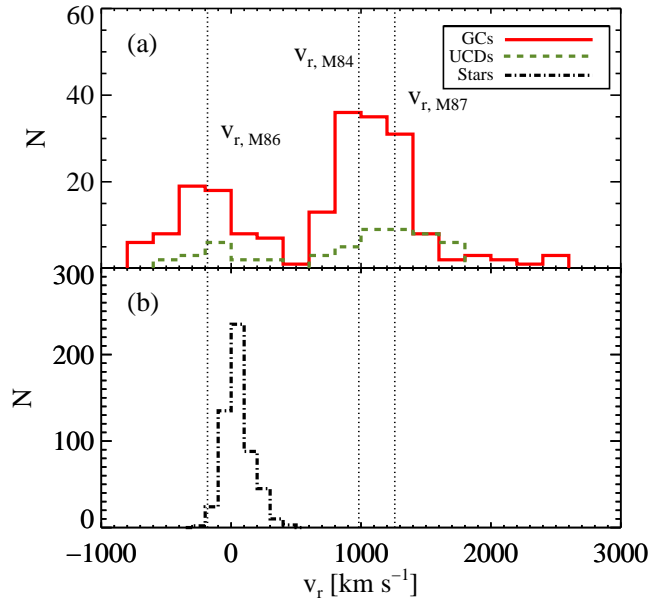}
\caption{(a) Radial velocity distribution of the GCs (red solid line) and the UCDs (dashed line) observed in this study.
(b) Radial velocity distribution of the foreground stars classified in this study. 
The dotted lines in two panels represent the systemic radial velocity of M86 (--182 km $\mathrm{s^{-1}}$), M84 (983 km $\mathrm{s^{-1}}$), and M87 (1260 km $\mathrm{s^{-1}}$) \citep{kim14}. 
\label{fig:v_dist}}
	\end{figure}

The radial velocity distributions of GCs and UCDs are shown in \textbf{Figure \ref{fig:v_dist}(a)}. 
The GCs and the UCDs have the radial velocities of --800 \kms to 2600 \kmsend. 
The radial velocity distribution of GCs shows two peaks at --200 \kms and 1100 \kmsend, which are similar to the radial velocities of M86 and M84, respectively. 
The UCDs show a similar radial velocity distribution to that of the GCs.
The radial velocity distribution of GCs with \vrad $>$ 400 \kms consists of at least two components, M87 and M84 GCs.
The peak at the radial velocity of M87 is not more prominent than that at the radial velocity of M84 because our survey region does not include the main body of M87.
These radial velocity distributions are distinct from that of the foreground stars that shows a strong concentration on \vrad $\sim$ 53 $\pm$ 5 \kms with a dispersion of 105 \kms (see \textbf{Figure \ref{fig:v_dist}(b)}).


	\subsubsection{Compilation of Velocity Data of GCs and UCDs}

We compile the velocity data of GCs and UCDs from this study and from the literature to study the kinematics of GCs and UCDs in the central region of the Virgo.
	\begin{figure}
\epsscale{1}
\includegraphics[trim=30 20 30 5,clip,width=\columnwidth]{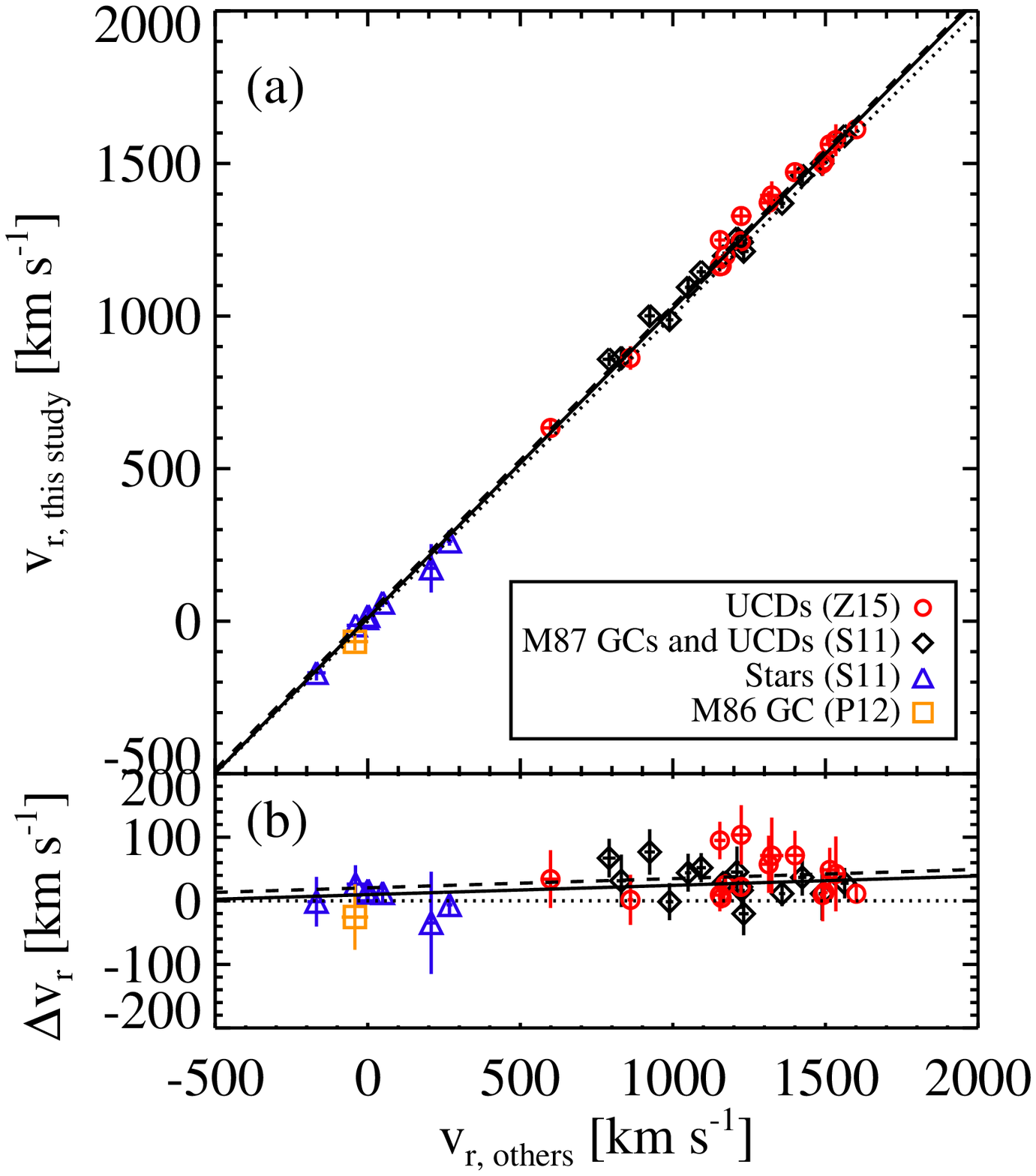}
\caption{(a) Comparison of radial velocities of the objects measured in this study with those in \citet{str11}, \citet{plh12}, and \citet{zha15}.
The circles, diamonds, triangles, and a square represent M87 UCDs \citep{zha15}, M87 GCs/UCDs, foreground stars near M87 \citep{str11}, and a M86 GC \citep{plh12}, respectively.
The dotted line denotes the one-to-one relation. The solid lines and dashed lines represent the least-square fit with the velocity data of \citet{str11} and that of \citet{zha15}, respectively. 
The horizontal and vertical error bars represent the measurement uncertainties of the radial velocities.
(b) Radial velocity difference $\mathrm{(v_{others} - v_{this ~study})}$. The symbols are the same as (a). \label{fig:v_comp}}	
	\end{figure}
Prior to the compilation, we compare the radial velocities of GCs, UCDs, and foreground stars derived in this study with the measurements from the literature (see \textbf{Figure \ref{fig:v_comp}}).
\citet{str11} presented the radial velocities of M87 GCs, UCDs, and foreground stars in the M87 field.
\citet{zha15} presented the radial velocities of M87 UCDs including all UCDs confirmed in \citet{str11}. 
In addition, \citet{plh12} presented those of M86 GCs. 
The numbers of common GCs and foreground stars in this study and \citet{str11} are 15 and 7, respectively, and the number of common UCDs in this study and \citet{zha15} is 16. For M86 GCs, there is only one common GC in both this study and \citet{plh12}.
\textbf{Figure \ref{fig:v_comp}} shows that our measurements agree well with previous measurements.

We fit to the data and derive a transformation relation between the measurements of this study and those of \citet{str11},
\begin{displaymath}
    v_\mathrm{r,this~study}=1.01~(\pm~0.007) \times v_\mathrm{r, S11}~+~9.69~(\pm~7.39)~\mathrm{km~s^{-1}}.
\end{displaymath}
Similarly, the transformation relation between the measurements of this study and those of \citet{zha15},
\begin{displaymath}
    v_\mathrm{r,this~study}=1.01~(\pm~0.02) \times v_\mathrm{r,Z15}~+~20.45~(\pm~27.53)~\mathrm{km~s^{-1}}.
\end{displaymath}

The root mean square errors of above two relations are about 25 and 32 \kmsend, respectively, which are comparable with the mean error of radial velocity estimates in this study (28 \kmsend). 
We transform the radial velocities of GCs and UCDs in \citet{str11} and \citet{zha15} using the above relations, and produce a master catalog of radial velocities for the GCs and UCDs in the central region of the Virgo.

	\begin{figure}
\epsscale{1}
\includegraphics[trim=10 20 0 25,clip,width=\columnwidth]{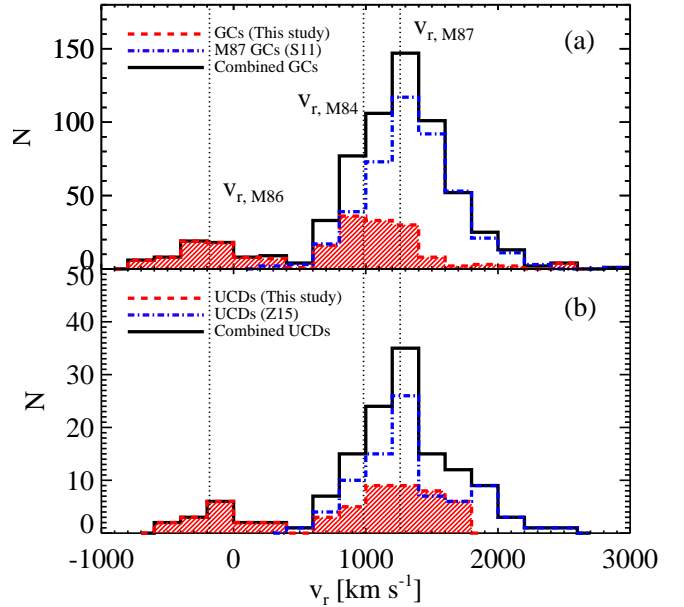}
\caption{(a) Radial velocity distribution of the GCs in the central region of the Virgo confirmed in this study (hatched histogram) and \citet{str11} (dot-dashed histogram).
The solid histogram indicates the radial velocity distribution for the combined GC sample.
(b) Same as (a), but for the UCDs. The dot-dashed histogram represents the UCDs presented in \citet{zha15}.
\label{fig:master_vdist}}	
	\end{figure}	
\textbf{Figure \ref{fig:master_vdist}(a) and (b)} show the radial velocity distributions for 633 GCs and 138 UCDs in the Virgo central region, respectively.
The radial velocity distributions of GCs and UCDs show two groups with peak positions of \vrad $\sim$ 1300 and --180 \kmsend.
The peak positions are consistent with the radial velocities of M87 and M86 (\vrad = 1260 \kms and --182 \kmsend). 
The GCs and UCDs with \vrad $>$ 400 \kms are expected to consist of at least two components including M87 and M84 populations.

To decompose spatially and kinematically different groups in the GCs, we perform three dimensional Gaussian mixture modeling for $\Delta$R.A., $\Delta$Dec., and radial velocities using the {\sc mclust} package in the {\sc R} statistical computing language \citep{fr02, fra12}.
The {\sc Mclust} finds the optimal number of components in a given sample set by calculating the Bayesian information criterion (BIC, Schwarz 1978) based on 13 different mixture models for multivariate data.
It detects 8 and 3 different groups in the master GC and UCD catalogs, respectively.
The model at which the optimal BIC occurs for the GC and UCD samples is the VVI (diagonal, varying volume and shape) model.

	\begin{figure*}
\epsscale{1}
\plotone{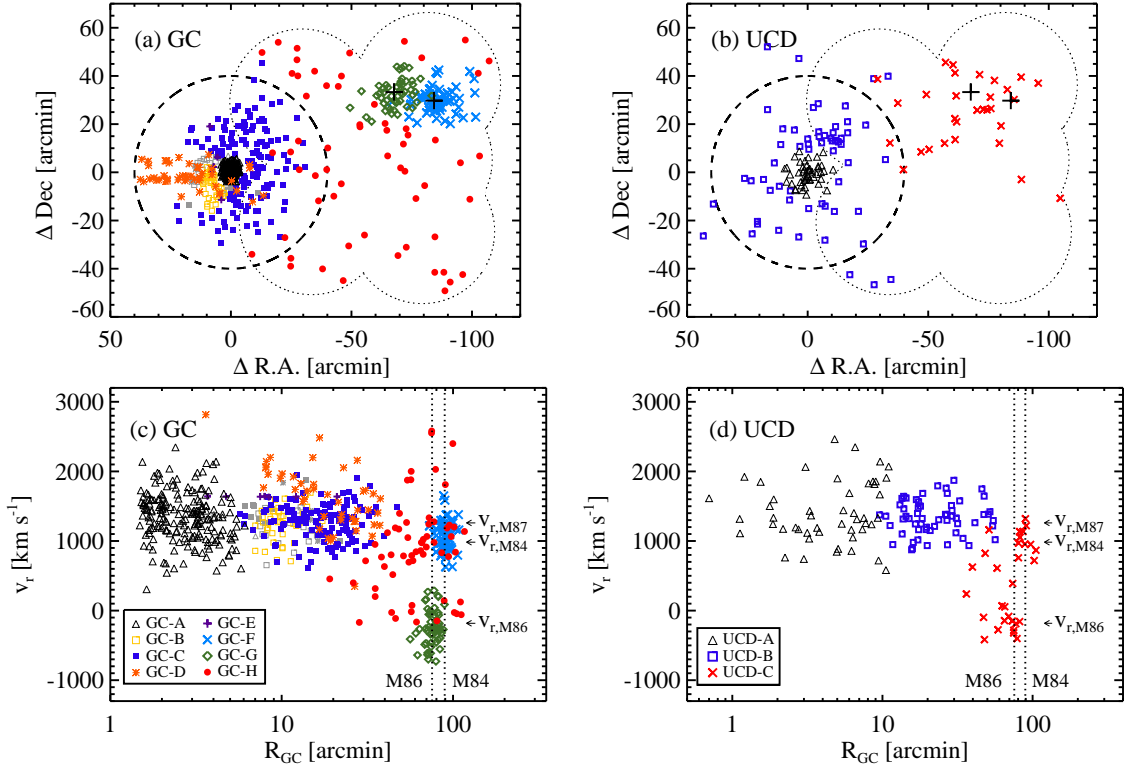}
\caption{(a) Spatial distribution of the master GC sample (this study and Strader et al. 2011).
(0,0) indicates the center of M87.
A dashed-line open circle centered on M87 has the radius of 40 arcmin where the slope of the GC number density profile start to change \citep{lph10}. 
The dotted outlines show the survey region of this study. 
The two large pluses represent the position of M86 and M84.
(b) Spatial distribution of the master UCD sample (this study and Zhang et al. 2015). The lines and pluses represent the same as (a).
(c) Radial velocity vs. galactocentric distance from M87 of the master GC sample.
The two dotted vertical lines represent the position of M86 and M84.
(d) Radial velocity vs. galactocentric distance from M87 of the master UCD sample. The lines represent the same as (c).
The symbols in four panels show different GCs belonging to different components as found by \textsc{Mclust}.
\label{fig:spatial}}
	\end{figure*}
{\bf Figure \ref{fig:spatial}} shows the spatial distribution and the radial velocity as a function of galactocentric distance from M87 for the master GC and UCD sample.
The GCs are divided into 8 different groups: GC-A, B, C, D, E, F, G, and H.
Most of the GC-A, B, C, D, and E groups are located within \rgc = $40\arcmin$ (see Figure \ref{fig:spatial}(a)), where a break is shown in the radial number density profile of GCs \citep{lph10}.
We consider 460 GCs in these five groups as the GCs that belong to M87.
The several groups for M87 GCs might be because of substructures in M87 \citep{str11, rom12} and of not uniform spatial coverage of observed data.
The GC-F and G groups are well concentrated on the position of M84 and M86, both spatially and kinematically, respectively (see Figures \ref{fig:spatial}(a) and (c)).
Therefore, the GC-F and G groups are considered as M84 and M86 GCs, respectively.
The GC-H group spreads out on the entire survey region, and is not associated with M84, M86, and M87. They could be IGC candidates. The selection method of IGCs in the Virgo will be described in \S3.1.3.

The UCDs are divided into three different groups: UCD-A, B, and C (see Figure \ref{fig:spatial}(b) and (d)).
Most of the UCD-A and B groups are located within the M87 region.
The UCD-C groups show a weaker concentration on M86 and M84.
Unlike the GC sample, the UCDs around M84 and M86 are not clearly distinguished from those in the intracluster region, because the number of the UCD sample is much smaller than that of the GC sample.
We divide the UCDs into two groups, M87 UCDs (UCD-A and B groups) and the others (UCD-C group), for the following analysis.
The 3D GMM results derived by {\sc Mclust} are summarized in {\bf Table \ref{tab:gc_ucd_sub}.}

\begin{deluxetable*}{c r c c c c c c c l}
\tablecaption{Summary of {\sc Mclust} Results$^{\rm a}$ for $\Delta$R.A., $\Delta$Dec. and radial velocities of the master GC and UCD Sample \label{tab:gc_ucd_sub}}
\tablewidth{0pt}
\tablehead{
\colhead{Group} & \colhead{N} & 
\colhead{Mean$_{\rm \Delta{R.A.}}$} & \colhead{$\sigma_{\rm \Delta{R.A.}}$} & 
\colhead{Mean$_{\rm \Delta{Dec.}}$} & \colhead{$\sigma_{\rm \Delta{Dec.}}$} & 
\colhead{Mean$_{\rm v_r}$} & \colhead{$\sigma_{\rm v_r}$} & \colhead{Fraction} & \colhead{Comment} \\
\colhead{} & \colhead{} & \colhead{[arcmin]} & \colhead{[arcmin]} & \colhead{[arcmin]} & \colhead{[arcmin]} & \colhead{[\kmsend]} & \colhead{[\kmsend]} & \colhead{} & \colhead{}
}
\startdata
GC-A & 202 & 0.0 & 1.8 & 0.9 & 2.4 & 1350 & 350 & 0.28 & M87 GCs \\
GC-B & 62 & 7.0 & 5.2 & --3.7 & 6.2 & 1281 & 259 & 0.09 & M87 GCs \\
GC-C & 138 & --4.1 & 11.2 & 3.2 & 14.6 & 1267 & 268 & 0.23 & M87 GCs \\
GC-D & 51 & 12.0 & 13.4 & --1.9 & 4.5 & 1481 & 411 & 0.11 & M87 GCs \\
GC-E & 7 & 5.5 & 2.4 & 3.3 & 8.5 & 1637 & 4 & 0.01 & M87 GCs \\
GC-F & 62 & --85.8 & 6.3 & 29.9 & 5.0 & 1067 & 202 & 0.09 & M84 GCs \\
GC-G & 51 & --68.5 & 6.0 & 32.8 & 5.5 & --247 & 252 & 0.08 & M86 GCs \\
GC-H & 60 & --54.3 & 29.0 & 6.8 & 30.2 & 916 & 680 & 0.11 & IGC candidates \\
\hline
UCD-A & 49 & 0.3 & 4.6 & --0.4 & 4.8 & 1424 & 440 & 0.32 & UCDs in the M87 region \\
UCD-B & 59 & --2.4 & 16.8 & 2.6 & 21.1 & 1328 & 257 & 0.46 & UCDs in the M87 region \\
UCD-C & 30 & --66.7 & 18.2 & 24.6 & 14.3 & 416 & 584 & 0.21 & 
\enddata

\tablenotetext{a}{Based on the VVI (diagonal, varying volume and shape) models for both GC and UCD sample, respectively.}

\end{deluxetable*}


		\subsubsection{Identification of Intracluster GCs}

We select the intracluster population from our GC sample based on the {\sc Mclust} results presented in the previous section.
The GC-H group is considered to be the IGC candidates not associated with M87, M86, and M84.
There are 60 IGC candidates in the GC-H group, consisting of 59 GC sample from this study and one GC (ID: H58443) from \citet{str11}.

In addition, we try a different method to select the IGCs in the GC sample confirmed in this study using three criteria.
First, IGCs are not located within \rgc $<$ 40$\arcmin$.
\citet{lph10} found a break in the radial number density profile of GCs at the galactocentric radius of 40$\arcmin$ from M87. 
We therefore use $R = 40\arcmin$ to distinguish M87 GCs from IGCs.
Second, IGCs are not located in any galaxy region defined by $R \sim 2D_{25}$.
This is a more strict criterion for selecting IGCs than the one used in Fornax Cluster ($R > 1.5D_{25}$; Bergond et al. 2007).
Third, when the GCs are located at $R < 2D_{25}$ of individual galaxies, but have radial velocities significantly different from the galaxies (i.e. $\Delta$\vrad $>$ 600 \kmsend), we consider them as IGCs.
These criteria result in 47 IGCs.

We compare the IGCs selected by two different methods described above.
We call the first method based on 3D GMM results as ``3D GMM method" and the second method with spatial and kinematic information as  ``Rv cut method".
There are 38 IGCs satisfying both two selection methods.
These are about 63\% and 83\% of the IGCs selected by the 3D GMM and Rv cut methods, respectively.
	\begin{figure}
\epsscale{1}
\includegraphics[trim=50 10 30 110,clip,width=\columnwidth]{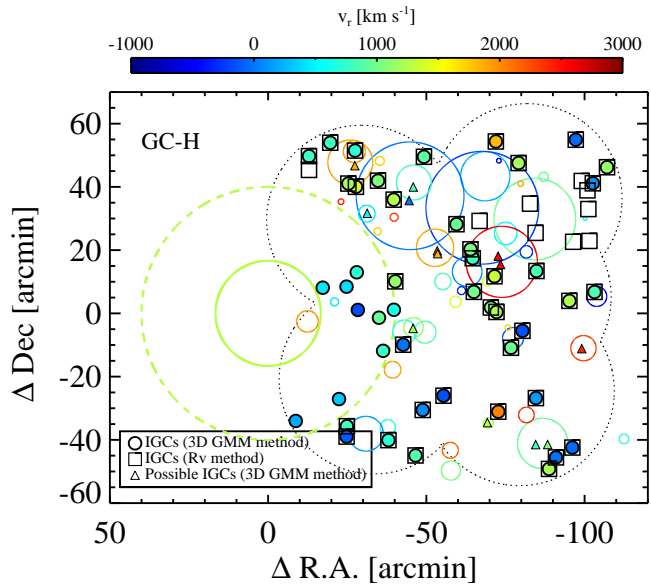}
\caption{Spatial distribution of the IGCs and possible IGCs color-coded by their radial velocities.
The filled circles and triangles represent the IGCs and possible IGCs based on the {\sc Mclust} results, respectively.
The open squares represent the IGCs selected by different criteria including R $> 2D_{25}$ and $\Delta$\vrad $< 600$ \kms (Rv method).
The detailed selection criteria is described in \S3.1.3.
The dashed line and the dotted outline are the same as Figure \ref{fig:spatial}.
The solid-line open circles indicate bright Virgo galaxies in the survey region color-coded by their radial velocities \citep{kim14}. 
The radius of each solid-line open circle represents 2$\mathrm{D_{25}}$ of each galaxy from RC3 catalog \citep{dev91}.
\label{fig:spatial_igc}}
	\end{figure}
{\bf Figure \ref{fig:spatial_igc}} shows the spatial distribution of these IGC candidates overlaid on the region of Virgo galaxies in the survey region.
The most prominent differences between the IGCs from two methods occurs at \rgc $< 40\arcmin$ and the western part of M84.
Using the 3D GMM method, 8 IGCs within 40$\arcmin$ from M87 are statistically well distinguished from M87 GCs.
Some IGCs around M84 selected by the Rv cut method are excluded in the IGCs from the 3D GMM method.

We adopt both methods to select the IGCs in the Virgo.
The 3D GMM method is more objective than the Rv method, but is not efficient for the decomposition of the group containing the small number of samples.
We apply the same spatial and radial velocity criteria of Rv method for 60 IGC candidates based on 3D GMM method, 
and find that 14 of the 60 GCs are associated with Virgo galaxies.
We define them as ``possible IGCs".
As a result, 46 and 14 are classified as IGCs and possible IGCs, respectively, among 60 IGC candidates based on 3D GMM method.
We use this 46 IGCs for the following analysis.

    \subsection{Radial Velocity and Velocity Dispersion Profiles of GCs and UCDs}


	\begin{figure}
\epsscale{1}
\includegraphics[trim=10 0 0 20,clip,width=\columnwidth]{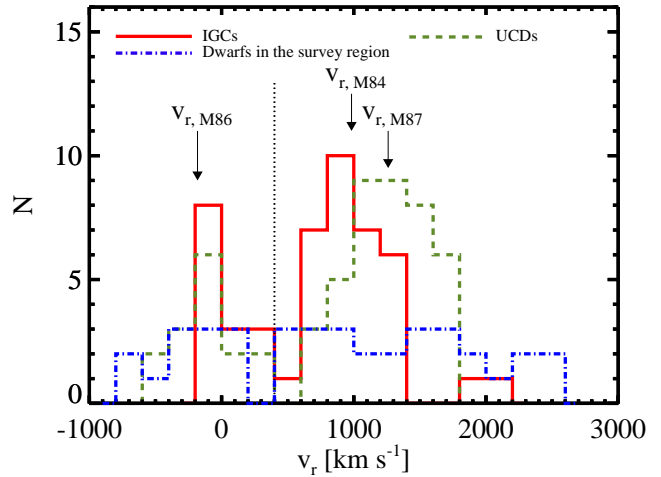}
\caption{Radial velocity distribution of IGCs confirmed in this study (solid histogram), UCDs confirmed in this study (dashed histogram), and dwarfs in the survey region \citep{kim14} (dot-dashed histogram).
\label{fig:igc_v}}
	\end{figure}	
	\begin{figure}
\epsscale{1}
\includegraphics[trim=10 5 25 20,clip,width=\columnwidth]{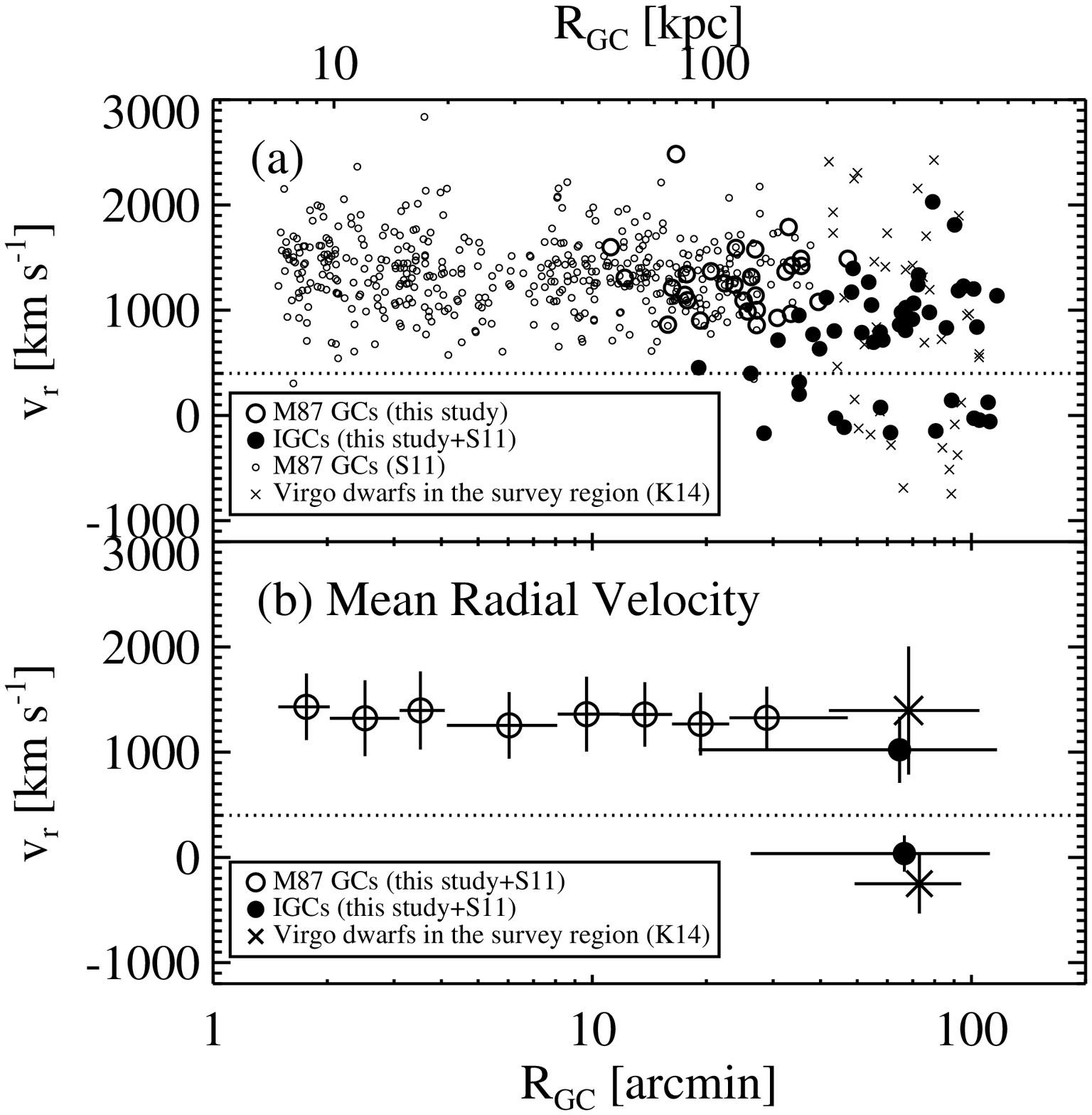}
\caption{(a) Radial velocities of GCs and dwarfs as a function of galactocentric distance from M87. 
The filled, open circles, and crosses represent the IGCs, M87 GCs, and dwarfs in the survey region \citep{kim14}, respectively.
The small and large open circles represent the M87 GCs in \citet{str11} and in this study, respectively.
The dashed horizontal line indicates \vrad = 400 \kmsend, where a dip in the radial velocity distribution exists  (see Figure \ref{fig:igc_v}).
(b) Mean radial velocity of GCs and dwarfs in each radial bin as a function of galactocentric distance from M87.
The symbols are the same as (a).
We divided the IGCs and dwarfs into two groups, respectively, according to the radial velocities, \vrad $>$ 400 \kms and with \vrad $<$ 400 \kmsend.
The vertical error bars indicate the root mean square velocities of the objects in each bin.
 \label{fig:radv_gc}}	
	\end{figure}
	\begin{figure}
\epsscale{1}
\includegraphics[trim=10 5 25 20,clip,width=\columnwidth]{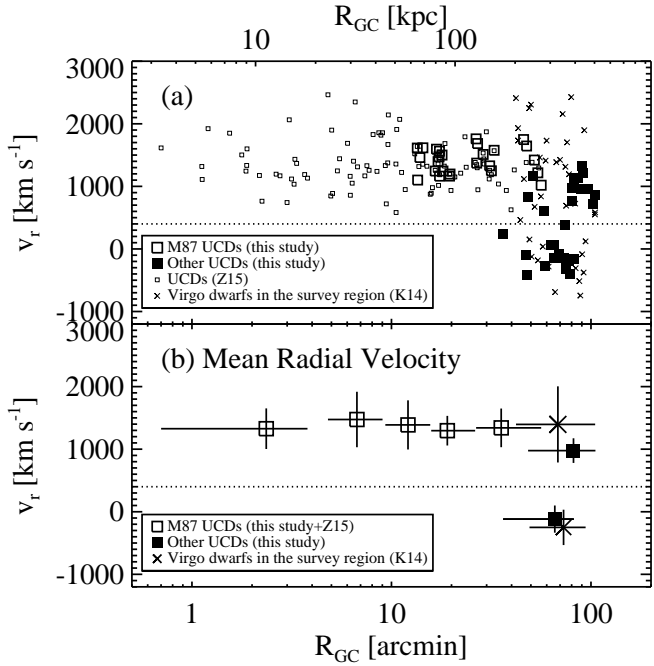}
\caption{(a) Radial velocities of UCDs and dwarf galaxies as a function of galactocentric distance from M87.
The open, filled squares, and crosses represent the UCDs in the M87 region, the UCDs not in the M87 region, and dwarfs in the survey region, respectively.
The small and large open squares represent the UCDs in \citet{zha15} and in this study, respectively.
(b) Mean radial velocity of UCDs and dwarfs in each radial bin as a function of galactocentric distance from M87. 
The symbols are the same as (a).
We divided the dwarfs into two groups, according to the radial velocities, \vrad $>$ 400 \kms and with \vrad $<$ 400 \kmsend.
The vertical error bars indicate the root mean square velocities of the objects in each bin.
\label{fig:radv_ucd}}	
	\end{figure}
	\begin{figure}
\epsscale{1}
\includegraphics[trim=10 5 25 0,clip,width=\columnwidth]{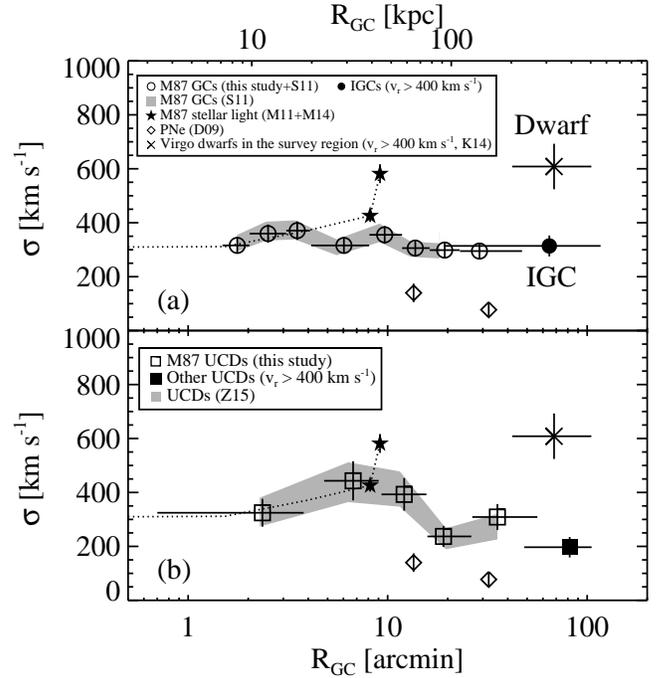}
\caption{Radial velocity dispersion profile of different kinds of objects in the Virgo as a function of galactocentric distance from M87.
The diamonds and filled star symbols with the dotted line represent the PNe \citep{doh09} and the stellar light of M87 (Murphy et al. 2011, 2014), respectively. 
The open circles, filled circle, open squares, filled squares, and cross represent M87 GCs, IGCs with \vrad $>$ 400 \kmsend, M87 UCDs, the other UCDs with \vrad $>$ 400 \kmsend, and dwarfs with \vrad $>$ 400 \kms in the survey region, respectively.
The vertical error bars represent the measurement uncertainties of the radial velocity dispersion calculated in each radial bin.
\label{fig:v_disp}}
	\end{figure}
We compare the radial velocity distribution of IGCs, UCDs, and dwarf galaxies in the survey region (see \textbf{Figure \ref{fig:igc_v}}).
The radial velocity distribution of the IGCs shows two prominent peaks at \vrad = --100 and 900 \kmsend.
The UCDs are also divided into two groups of which radial velocities are at \vrad = --100 and 1100 \kmsend.
The peak position of the radial velocity distribution of the UCDs with \vrad $>$ 400 \kms is higher than that of the IGCs with \vrad $>$ 400 \kms because there are the UCDs in the M87 region (UCD-A and B groups, see \S3.1.2).
The radial velocities of the dwarfs in the survey region are almost uniformly distributed, different with the IGCs and UCDs.
However, there is a dip at \vrad $\sim$ 400 \kms in the radial velocity distribution of the dwarfs in the survey region, which is also shown in those of IGCs and UCDs.
We assume that the objects with \vrad $>$ 400 \kms belong to the main body of the Virgo Cluster, while the others are infalling or outgoing populations in the Virgo.
Therefore, we divide the IGCs, UCDs, and dwarfs into two groups in each population according to their radial velocities in order to discuss their kinematics separately.

\textbf{Figure \ref{fig:radv_gc}(a)} shows the radial velocities of the combined GC sample and dwarfs in the survey region as a function of galactocentric distance from M87.
The dwarfs in the survey region have the radial velocities with a wide range between --745 \kms and 2426 \kmsend.
We also divide the dwarfs into two categories, one with \vrad $>$ 400 \kms and the others with \vrad $<$ 400 \kmsend.
\textbf{Figure \ref{fig:radv_gc}(b)} shows the mean radial velocity of M87 GCs, IGCs, and dwarfs in each radial bin as a function of galactocentric distance from M87. 
The mean radial velocities of the M87 GCs with \vrad $>$ 400 \kms are almost contant ($\overline{v_{\rm r}}$ = 1340 \kmsend) within their uncertainties.
For the IGCs and dwarfs, we calculate the mean radial velocities of two groups with high and low radial velocities, respectively.
The mean radial velocity of the IGCs with \vrad $>$ 400 \kms ($\overline{v_{\rm r}}$ = 1023 \kmsend) is lower than those of dwarfs with \vrad $>$ 400 \kms ($\overline{v_{\rm r}}$ = 1396 \kmsend) and M87 GCs.
The mean radial velocities of the IGCs and dwarfs with \vrad $<$ 400 \kms are 36 and --250 \kmsend, respectively.
\textbf{Figure \ref{fig:radv_ucd}} is similar to \textbf{Figure \ref{fig:radv_gc}}, but for the combined UCD sample.
The mean radial velocity of the M87 UCDs is almost constant ($\overline{v_{\rm r}}$ = 1365 \kmsend).
The other UCDs (UCD-C group, see \S3.1.2) include two groups with \vrad $>$ 400 \kms and \vrad $<$ 400 \kmsend.
The mean radial velocity of the other UCDs with \vrad $>$ 400 \kms ($\overline{v_{\rm r}}$ = 977 \kmsend) is lower than those of dwarfs with \vrad $>$ 400 \kms and M87 UCDs.
The mean radial velocity of the other UCDs with \vrad $<$ 400 \kms is --117 \kmsend.

\textbf{Figure \ref{fig:v_disp}} shows the radial velocity dispersion profiles of GCs and UCDs.
The radial velocity dispersion and its error of the GCs, UCDs, and dwarfs in each radial bin are calculated following the formula in \citet{pm93}.
We use the IGCs, UCDs, and dwarfs with \vrad $>$ 400 \kms that are not probably affected by infalling populations to derive velocity dispersion for each tracer.
The radial velocity dispersion of the combined M87 GCs is almost constant at 333 \kmsend.
The radial velocity dispersion of the IGCs is similar to that of the GCs in M87 ($\sigma$ = 314 \kmsend).
They are much lower than the radial velocity dispersion of dwarfs in the same survey region ($\sigma$ = 608 \kmsend) and that of ICPNe superposed on the M87 region (Longobardi et al. 2015; $\sigma$ = 881 \kmsend).
\citet{str11} suggested that the stellar halo of M87 is not truncated at \rgc $\sim 30\arcmin$, which is different from the result from the PNe of M87 \citep{doh09}.
Our results are consistent with \citet{str11}; the dispersion profile of GCs does not show any significant changes with galactocentric radius.
The radial velocity dispersion profile of the combined UCDs is similar to that of the UCDs in \citet{zha15}.
Similar to the case of the GCs, the dispersion profile of UCDs does not change much at \rgc $> 40\arcmin$, either.
The kinematic information of GCs and UCDs is summarized in \textbf{Tables \ref{tab:gc_kinematics} and \ref{tab:ucd_kinematics}}, respectively.

As shown in \textbf{Figure \ref{fig:v_disp}}, the velocity dispersion profiles of GCs, UCDs, the integrated stellar light, and PNe are remarkably different from each other.
The dispersion profile differs significantly at \rgc = $7\arcmin-40\arcmin$.
This is where the cD halo of M87 may exist \citep{mih05,kor09}.
The GC velocity dispersion is almost constant at $\sigma \sim$ 300 -- 350 \kms in this range, 
while the stellar velocity dispersion increases steeply and the PNe velocity dispersion is much lower than the others. 
Note that most of GC samples in this range come from \citet{str11}, because our observation field is concentrated on the intracluster region (i.e. \rgc = $40\arcmin-120\arcmin$). 
This disagreement of the velocity dispersion has been already reported in the  previous studies (e.g. Romanowsky \& Kochanek 2001; Murphy et al. 2014). 
\citet{mgc14} concluded that the kinematic properties of different objects are not necessarily the same because the Virgo Cluster has not dynamically relaxed yet and because there are several substructures.
These diverse kinematic properties of various tracers in galaxy clusters are also found in other galaxy clusters (e.g. Richtler et al. 2011; Coccato et al. 2013).
A detailed dynamical analysis of different tracers by considering their number density profiles, velocity anisotropies, and the galaxy cluster mass profile is necessary to understand the kinematic difference.

\section{DISCUSSION}

    \subsection{The Number of Virgo IGCs}

The number density of IGCs in the Virgo has been predicted from several photometric studies that assumed a Gaussian form of the GC luminosity funciton.
\citet{tam06} suggested the number density of IGCs as $\sim$ 0.2 arcmin$^{-2}$, ranging from 0.1 to 0.5 arcmin$^{-2}$.
This value is comparable with the result of \citet{dur14}.
\citet{har09}, \citet{str11}, and \citet{oa16} presented the background number density as 0.6, 0.51, and 0.58 arcmin$^{-2}$, respectively, that are slightly higher than above results.
These background objects may include foreground stars, background galaxies, and IGCs, which could not be distinguished with the photometric data only.
Therefore, these numbers are probably overestimated values.

\begin{deluxetable}{c c c c c}
\tablecaption{GC Kinematics \label{tab:gc_kinematics}}
\tablewidth{0pt}
\tablehead{
\colhead{R [arcmin]} & \colhead{$\overline{\rm{R}}$ [arcmin]} & \colhead{N} & \colhead{$\overline{v_\mathrm{r}}$ [\kmsend]} & \colhead{$\sigma$ [\kmsend]}
}
\startdata
\multicolumn{5}{c}{M87 GCs (this study + S11)}\\
\hline
 1.49 --  2.03 &  1.76 & 58 & 1431 $\pm$ 41 & 316 $\pm$ 30 \\
 2.03 --  3.10 &  2.52 & 58 & 1322 $\pm$ 47 & 360 $\pm$ 34 \\
 3.10 --  4.08 &  3.52 & 58 & 1396 $\pm$ 49 & 371 $\pm$ 35 \\
 4.14 --  8.08 &  6.04 & 58 & 1255 $\pm$ 41 & 316 $\pm$ 30 \\
 8.11 -- 11.81 &  9.66 & 58 & 1361 $\pm$ 47 & 355 $\pm$ 33 \\
11.81 -- 16.24 & 13.76 & 58 & 1358 $\pm$ 40 & 306 $\pm$ 29 \\
16.24 -- 22.95 & 19.31 & 58 & 1268 $\pm$ 39 & 299 $\pm$ 28 \\
23.03 -- 47.18 & 28.86 & 54 & 1327 $\pm$ 40 & 295 $\pm$ 29 \\
\hline
\multicolumn{5}{c}{All M87 GCs (this study + S11)}\\
\hline
 1.49 -- 47.18 & 10.52 & 460 & 1340 $\pm$ 16 & 333 $\pm$ 11 \\
\hline
\multicolumn{5}{c}{IGCs with \vrad $>$ 400 \kms (this study + S11)}\\
\hline
19.08 -- 116.79 &  64.62 & 33 & 1023 $\pm$ 55 & 314 $\pm$ 39 \\
\hline
\multicolumn{5}{c}{IGCs with \vrad $<$ 400 \kms (this study)}\\
\hline
26.19 -- 111.81 &  66.56 & 14 &   36 $\pm$ 46 & 172 $\pm$ 33 \\
\hline
\multicolumn{5}{c}{IGCs regardless \vrad (this study)}\\
\hline
19.08 -- 116.79 &  65.20 & 47 &  729 $\pm$ 77 & 530 $\pm$ 55
\enddata
\end{deluxetable}

\begin{deluxetable}{c c c c c}
\tablecaption{UCD Kinematics \label{tab:ucd_kinematics}}
\tablewidth{0pt}
\tablehead{
\colhead{R [arcmin]} & \colhead{$\overline{\rm{R}}$ [arcmin]} & \colhead{N} & \colhead{$\overline{v_\mathrm{r}}$ [\kmsend]} & \colhead{$\sigma$ [\kmsend]}
}
\startdata
\multicolumn{5}{c}{UCDs in the M87 region (this study + Z15)}\\
\hline
 0.70 --   3.79 &   2.36 & 21 & 1327 $\pm$ 70 & 324 $\pm$ 51 \\
 4.80 --   9.02 &   6.71 & 21 & 1474 $\pm$ 96 & 443 $\pm$ 72 \\
 9.32 --  15.64 &  12.10 & 21 & 1387 $\pm$ 85 & 392 $\pm$ 60 \\
15.83 --  26.27 &  19.05 & 21 & 1295 $\pm$ 51 & 236 $\pm$ 38 \\
26.56 --  56.29 &  35.49 & 21 & 1340 $\pm$ 67 & 308 $\pm$ 48 \\
\hline
\multicolumn{5}{c}{All UCDs in the M87 region (this study + Z15)}\\
\hline
 0.70 --  56.29 &  15.14 & 105 & 1365 $\pm$ 34 & 354 $\pm$ 25 \\
\hline
\multicolumn{5}{c}{UCDs not in the M87 region with \vrad $>$ 400 \kms (this study)}\\
\hline
48.40 -- 105.30 &  81.75 & 14 & 977 $\pm$ 52 & 196 $\pm$ 37 \\
\hline
\multicolumn{5}{c}{UCDs not in the M87 region with \vrad $<$ 400 \kms (this study)}\\
\hline
36.28 --  82.06 &  65.89 & 15 & -117 $\pm$ 56 & 217 $\pm$ 41
\enddata
\end{deluxetable}
	\begin{figure}
\epsscale{1}
\includegraphics[trim=10 10 10 10,clip,width=\columnwidth]{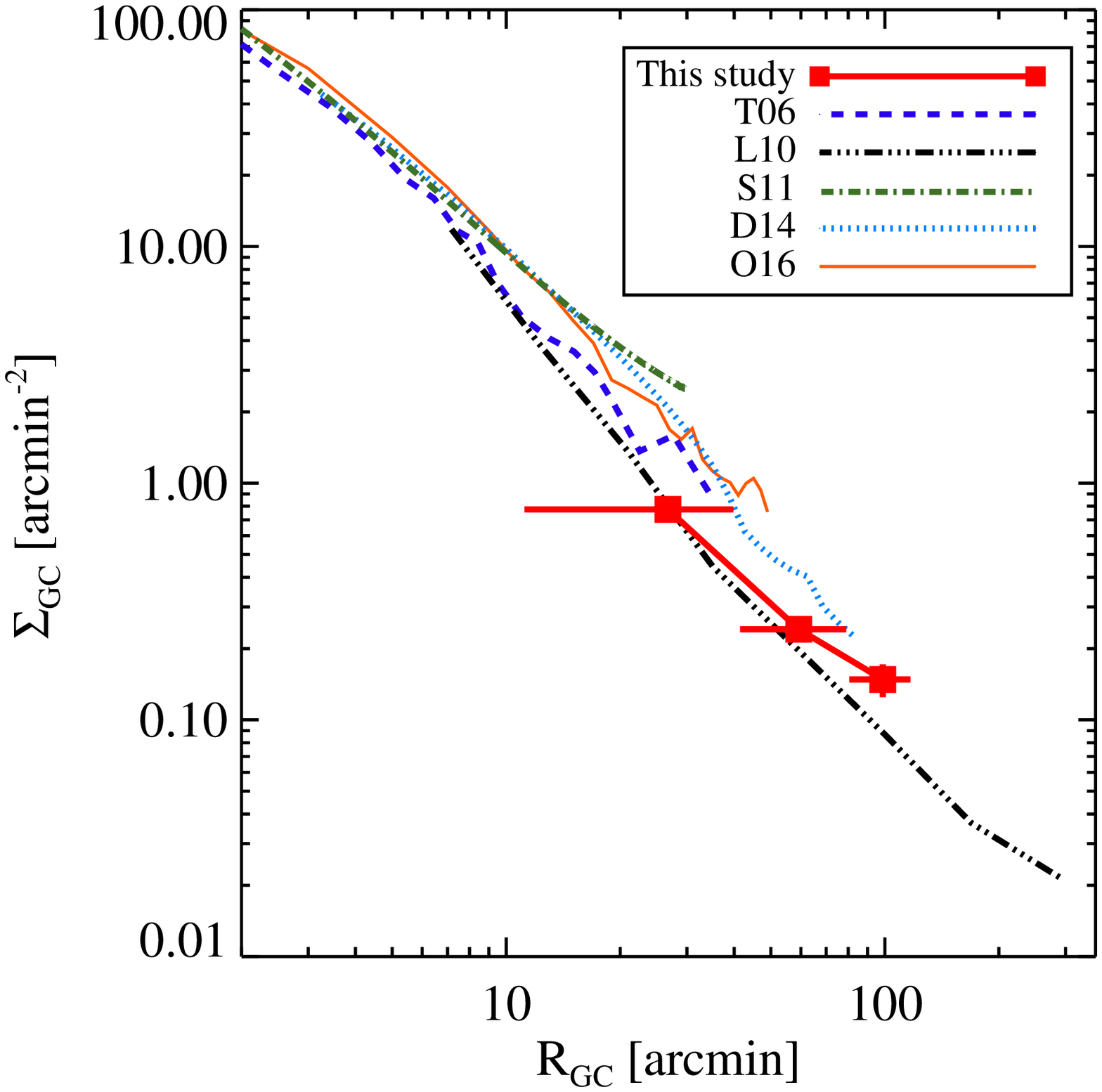}
\caption{Number density of M87 GCs and IGCs estimated within three radial bins (red filled squares) with radial number density profile of GCs in the Virgo derived by several references: \citet{tam06} (blue dashed line), \citet{lph10} (black dot-dot-dashed line), \citet{str11} (green dot-dashed line), \citet{dur14} (skyblue dotted line), and \citet{oa16} (orange solid line).
Note that the GCs within R $< 2.5\rm{D_{25}}$ for Virgo galaxies are excluded in the profile derived in \citet{lph10}.
\label{fig:rd}}
	\end{figure}
\textbf{Figure \ref{fig:rd}} shows the radial number density profiles of M87 GCs and Virgo GCs derived from various surveys \citep{tam06, lph10, str11, dur14, oa16}.
The number density profiles of \citet{str11}, \citet{dur14}, and \citet{oa16} are similar 
in the inner region at \rgc$< 20\arcmin$, but show some differences in the outer regions.
However, the number density profiles of \citet{tam06} and \citet{lph10} at \rgc $> 10\arcmin$ are steeper than the others.
This difference seems to mainly come from different spatial coverages in the two studies.
\citet{tam06} only covered the eastern part of M87 where there are no massive galaxies except M89, while
\citet{lph10} masked out the circular region ($R < 2.5D_{25}$) for Virgo galaxies to exclude the GCs that belong to the Virgo galaxies.

In order to estimate the total number of the IGCs, we assume that the luminosity function of the IGCs follows a Gaussian distribution with a peak of $g_0 =$ 23.8 $\pm$ 0.2 mag and a width of 1.4 \citep{dur14}.
In addition, the spectroscopic targets we selected are one third of the entire point sources with the same range of color and magnitude of GCs.
Considering these factors, we estimate the total number and the number density of IGCs in our survey region as about 2114 and 0.19 $\pm$ 0.02 arcmin$^{-2}$, respectively.
This is comparable with the estimates from photometric studies of $\Sigma_\mathrm{GC} \sim 0.2-0.6$ arcmin$^{-2}$ (see \textbf{Figure \ref{fig:rd}}).

Moreover, \citet{str11} predicted that 16-67\% of the IGCs have extreme radial velocities of \vrad $<$ 307 \kms or \vrad $>$ 2307 \kmsend.
In our sample, there are 12 IGCs with \vrad $<$ 307 \kms and no IGCs with \vrad $>$ 2307 \kmsend.
Among the 37 dwarf galaxies in the survey region, there are 12 and 2 dwarfs with low and high extreme velocities.
Therefore, the fractions of IGCs and dwarfs with extreme velocities are, respectively, 26\% and 38\%, consistent with the expectation of \citet{str11}.

	\subsection{Color Distribution of IGCs}

The GCs in massive early-type galaxies are generally divided into two subpopulations, blue GCs and red GCs (Brodie \& Strader 2006 and reference therein).
The colors of the old GCs depend mainly on their metallicity, the blue and red GCs are considered as metal-poor and metal-rich populations, respectively.
Several photometric surveys of M87 GCs reported that M87 also has these two different GC populations \citep{pen06, tam06, har09, oa16}.

	\begin{figure}
\epsscale{1}
\includegraphics[trim=20 0 20 10,clip,width=\columnwidth]{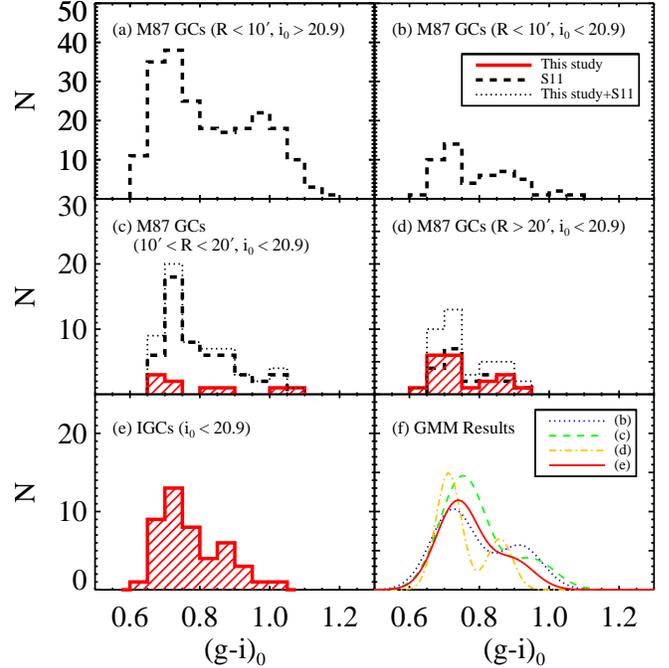}
\caption{Color distributions for M87 GCs and IGCs within several radial bins. (a) M87 GCs with \rgc $< 10\arcmin$ and $i_0 > 20.9$, (b) M87 GCs with \rgc $< 10\arcmin$ and $i_0 < 20.9$, (c) M87 GCs with $10\arcmin <$ \rgc $< 20 \arcmin$ and $i_0 < 20.9$, (d) M87 GCs with \rgc $> 20\arcmin$ and $i_0 < 20.9$, (e) IGCs with $i_0 < 20.9$, and (f) the GMM results for the color distribution of GCs in each radial bin adopting a homoscedastic case (same variances).
The solid, dashed, and dotted lines in (a)-(e) panels represent the GCs confirmed in this study, the GCs presented in \citep{str11}, and the combined GC sample, respectively.
\label{fig:rbin_color}}
	\end{figure}
\textbf{Figure \ref{fig:rbin_color}} shows the color distributions of spectroscopically confirmed GCs in this study and in \citet{str11}.
The $gi$ magnitudes of \citet{str11} based on the SDSS system are converted to CFHT AB magnitude system.
We divided the GC sample of \citet{str11} into two groups according to the GC magnitudes,
bright GCs with $i_0 < 20.9$ and faint GCs with $i_0 > 20.9$, because their sample contains fainter GCs than our sample.
All of the M87 GCs within each radial bin show a bimodal color distribution as expected by photometric surveys (see \textbf{Figure \ref{fig:rbin_color}(a)-(d)}).
It is noted that the color distribution of the IGCs confirmed in this study is not unimodal (see \textbf{Figure \ref{fig:rbin_color}(e)}).
We use the GMM adopting a homoscedastic case (same variances) to decompose these color distributions of the GCs within each radial bin into two components, blue and red ones. 
The GMM results are listed in \textbf{Table \ref{tab:color_dist}}, and are plotted in \textbf{Figure \ref{fig:rbin_color}(f)}.
\begin{deluxetable*}{c c c c c c c c c c l}
\tablecaption{Summary of GMM Tests$^\mathrm{a}$ for the GC Color Distribution \label{tab:color_dist}}
\tablewidth{0pt}
\tablehead{
\colhead{Region} & \colhead{$(g-i)_{0,\rm{blue}}$} & \colhead{$\sigma_{\rm{blue}}$} & \colhead{$f_{\rm{blue}}$} & \colhead{$(g-i)_{0,\rm{red}}$} & \colhead{$\sigma_{\rm{red}}$} & \colhead{$f_{\rm{red}}$} & \colhead{$N_{\rm{GC}}$} & \colhead{$p$} & \colhead{$D$}
}
\startdata
\multicolumn{9}{l}{Faint M87 GCs ($i_0 \ge 20.9$)} \\
\hline
 R $<$ 10$\arcmin$ & 0.74 $\pm$ 0.01 & 0.07 $\pm <$0.01 & 0.61 $\pm$ 0.04 & 0.98 $\pm$ 0.01 & 0.07 $\pm <$0.01 & 0.39 $\pm$ 0.04 & 216 & 2.33e-11 & 3.37 $\pm$ 0.25 \\
\hline
\multicolumn{9}{l}{Bright M87 GCs ($i_0<20.9$)} \\
\hline
 R $<$ 10$\arcmin$ & 0.72 $\pm$ 0.02 & 0.06 $\pm$0.01 & 0.65 $\pm$ 0.10 & 0.92 $\pm$ 0.03 & 0.06 $\pm$ 0.01 & 0.35 $\pm$ 0.10 & 51 & 9.78e-03 & 3.01 $\pm$ 0.66 \\
 10$\arcmin <$ R $<$ 20$\arcmin$ & 0.75 $\pm$ 0.01 & 0.07 $\pm$ 0.01 & 0.79 $\pm$ 0.08 & 0.95 $\pm$ 0.03 & 0.07 $\pm$ 0.01 & 0.21 $\pm$ 0.08 & 61 & 1.14e-03 & 3.00 $\pm$ 0.59 \\
 R $>$ 20$\arcmin$ & 0.71 $\pm$ 0.01 & 0.04 $\pm$ 0.01 & 0.70 $\pm$ 0.07 & 0.86 $\pm$ 0.01 & 0.04 $\pm$ 0.01 & 0.30 $\pm$ 0.07 & 39 & 3.56e-04 & 4.08 $\pm$ 0.70 \\
\hline
\multicolumn{9}{l}{Bright IGCs ($i_0<20.9$)} \\
\hline
 - & 0.74 $\pm$ 0.02 & 0.06 $\pm$ 0.01 & 0.75 $\pm$ 0.13 & 0.90 $\pm$ 0.05 & 0.06 $\pm$ 0.01 & 0.25 $\pm$ 0.13 & 45 & 6.03e-02 & 2.59 $\pm$ 0.74
\enddata
\tablenotetext{a}{Based on the homoscedastic (same variances) option. $p$ represents the probability for unimodality and $D$ is the difference in color between the two peaks. $f$ indicates the number ratio of each component.}
\end{deluxetable*}
We use the probability for the unimodal distribution ($p$) and the peak separation relative to the widths ($D$) as the indicators of bimodality (e.g. $D > 2$ for the clear separation of two peaks).
The GMM results show that the color distributions of M87 GCs at \rgc $< 10\arcmin$, $10\arcmin <$ \rgc $<20\arcmin$, and $20\arcmin <$ \rgc $< 40\arcmin$ are not probably unimodal but bimodal ($p < 0.01$ and $D > 2$). 
The IGC sample shows $p \sim 6\%$ and $D = 2.59 \pm 0.74$.
This means that the IGCs has $\sim$ 6\% probability of the unimodal color distribution, and the color distribution of the IGCs marginally shows a peak separation because of a large error for the $D$ value.
The blue and red peaks for the IGCs separated from GMM are similar to those for M87 GCs (see {\bf Table \ref{tab:color_dist}}).
From this we conclude that there exist red IGC populations even though the color distribution of the IGCs does not show a clear bimodality.

In \textbf{Figure \ref{fig:rbin_color}(f)}, the faint GCs ($i_0 > 20.9$) within \rgc $< 10\arcmin$ have blue and red peaks at redder colors than the bright GCs ($i_0 < 20.9$) in the same region.
The color difference of the red peaks is more significant than that of the blue peaks.
This trend is already reported by \citet{har09}.
On the other hand, the blue and red peaks for bright M87 GCs and IGCs are similar to each other within uncertainties except for the GCs with \rgc $> 20\arcmin$.

Many studies suggested that IGCs are stripped off from low-mass dwarf galaxies in galaxy clusters \citep{wil07, bek08, geo09, lee10}.
According to this scenario, most of IGCs are expected to be metal-poor compared with the GCs hosted by massive early-type galaxies.
However, we detect red IGCs of which the number ratio is about 24\% (see Table \ref{tab:color_dist}).
The red IGC population is also detected in the Coma Cluster \citep{pen11}.
The number of the red IGCs in the Coma is about 20\% of the number of the entire IGC population,
while the number ratio of red GCs of NGC 4874, a central massive galaxy of the Coma, is about 50\%.
\citet{pen11} suggested that the red IGCs are stripped from more massive galaxies rather than dwarf galaxies.
We also suspect that the red IGCs confirmed spectroscopically in this study originate from massive galaxies capable of hosting red GCs.

    \subsection{Dynamical State of the Virgo Cluster}
    
The Virgo Cluster consists of several different subclusters \citep{bst85, bts87, bpt93}: the main cluster A centered on M87, the cluster B centered on M49, three small clouds named as M, W, and W$\arcmin$ clouds \citep{dev61}, NGC 4636 group, and so on.
In the cluster A, there are two distinct substructures associated with M87 and M86, respectively.
Several studies suggested that M86 has its own group apart from the main cluster group centered on M87 \citep{bpt93, boh94, sbb99}.
The M86 group consists of M86 with \vrad = --181 \kmsend, NGC 4438 with \vrad = 10 \kmsend, and NGC 4402 with \vrad = 231 \kmsend.
The map of X-ray gas surrounding these galaxies suggests that these galaxies experience ongoing interaction \citep{ehl13}.
The radial velocity of the M86 group (\vrad $\sim$ --181 \kmsend) is $\sim$ 1400 \kms smaller than that of M87 (\vrad = 1260 \kmsend), suggesting that the M86 group may be infalling into the main cluster of the Virgo from behind.
\citet{jbb04} found a bimodality in the distance distribution of the galaxies in the cluster A, and suggested that the M86 group is falling into the main cluster from 1 Mpc behind, and \citet{mei07} confirmed this using the surface brightness fluctuation distances of galaxies.

	\begin{figure*}
\epsscale{1}
\plotone{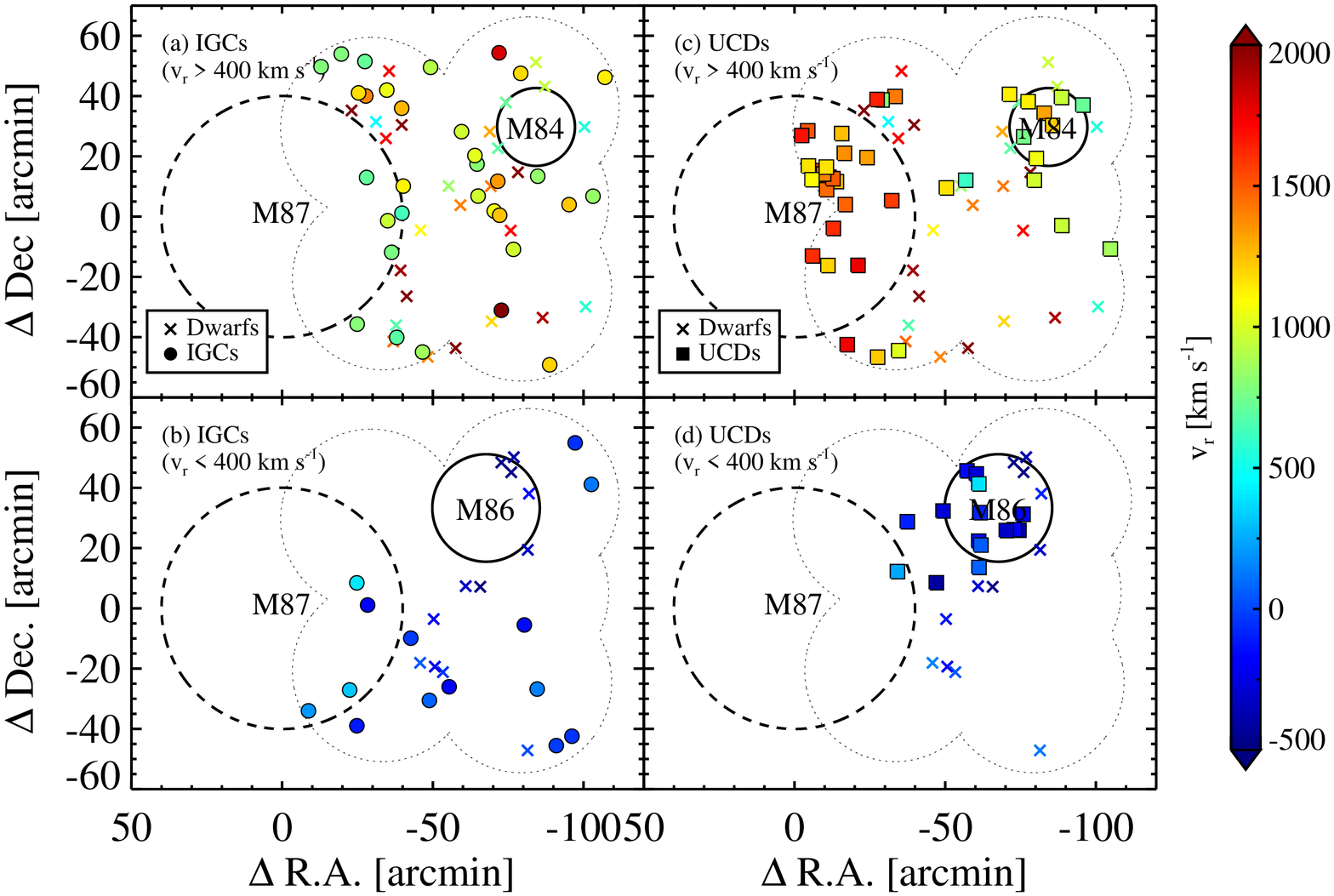}
\caption{(a) Spatial distribution of the IGCs with \vrad $>$ 400 \kms (circles) and dwarfs with \vrad $>$ 400 \kms (crosses).
(b) The same as (a), but for objects with \vrad $<$ 400 \kmsend.
(c) Spatial distribution of the UCDs with \vrad $>$ 400 \kms (squares) and dwarfs with \vrad $>$ 400 \kms (crosses).
(d) The same as (c), but for objects with \vrad $<$ 400 \kmsend.
The dashed-line open circle centered on M87 and dotted outline indicate the same as Figure \ref{fig:spatial}.
The solid-line open circles shows the galaxy region of M84 and M86 of which radii are 2$D_{25}$.
\label{fig:spatial_v}}
	\end{figure*}
\textbf{Figure \ref{fig:spatial_v}(a) and (c)} show the spatial distribution of the IGCs and UCDs with \vrad $>$ 400 \kms that probably belong to the cluster A of the Virgo, respectively.
Most of the IGCs spread out the north-western region of M87, while most of the UCDs are located around M87 and M84.
Few IGCs and UCDs with \vrad $>$ 400 \kms have the radial velocities greater than 2000 \kmsend, compared with the dwarfs.
Consequently, the radial velocity dispersions of the IGCs and UCDs with \vrad $>$ 400 \kms ($\sigma = 314 \pm 39$ \kms and $361 \pm 24$ \kmsend) are much lower than that of the dwarfs with \vrad $>$ 400 \kms in the survey region ($\sigma = 608 \pm 84$ \kmsend) (see \textbf{Figure \ref{fig:v_disp}}).
This indicates that the IGCs and UCDs are more dynamically relaxed than the dwarfs.
The strong peculiar motions of dwarfs might increase their velocity dispersion.
Therefore the mass of the Virgo based on dwarf galaxy kinematics may be an overestimate. 
Detailed dynamical analysis with Virgo cluster mass profiles and surface number density profiles of each tracer is necessary to understand the difference in the velocity dispersion between the dwarf galaxies and IGC/UCDs.

On the other hand, the velocity dispersion of the IGCs and UCDs with \vrad $<$ 400 \kms ($\sigma = 172 \pm 33$ \kms and $217 \pm 41$ \kmsend) is similar to that of the dwarfs with \vrad $<$ 400 \kms ($\sigma = 282 \pm 57$ \kmsend) within uncertainties.
The UCDs with \vrad $<$ 400 \kms are concentrated on the M86 region as well as most of dwarfs with \vrad $<$ 400 \kmsend, 
while the IGCs with \vrad $<$ 400 \kms are not (see \textbf{Figure \ref{fig:spatial_v}(b) and (d)}).
From this we consider that the UCDs and dwarfs with \vrad $<$ 400 \kms are related with the infalling group centered on M86.

	\subsection{The Origin of Virgo UCDs}

UCDs are very interesting objects of which formation scenario is still widely debated.
There are two representative scenarios for the origin of the UCDs: ``star cluster" and ``dwarf galaxy" origins.
Some studies suggested that they are simply the bright end of GCs \citep{hir99, mhi02, mhm12} or massive star clusters that were formed during star cluster mergers \citep{kro98, fk02, mar04, kjb06}.
Other studies proposed that they are the remnant nuclei of dwarf galaxies threshed by tidal interaction in the dense environment \citep{bmr94, bcd01, bek03, dri03, goe08, plj10, pb13, set14, jan15}.

There have been several observational studies on the Virgo UCDs to understand the formation mechanisms of the UCDs, 
and most of these studies supported the ``dwarf galaxy" origin.
\citet{has05} and \citet{jon06} found three and nine UCDs in the Virgo, respectively.
\citet{has05} suggest that the Virgo UCDs are the stripped nuclei of dwarf galaxies because of their mass-to-light ratios, luminosities, and colors similar to those of nucleated dwarfs, 
while \citet{jon06} did not make a concrete conclusion about the origin of UCDs.
\citet{bro11} complied a larger sample of Virgo UCDs and found that the 34 Virgo UCDs have the color-magnitude trend similar to that of the nuclei of dwarf galaxies and the kinematics distinct from that of normal GCs.
Recently, \citet{liu15} selected UCDs in the NGVS images and divided them into two groups: one with the visible envelope and another without any  visible envelope.
They found that the UCDs without the envelope are more centrally concentrated on M87 than those with envelopes, which are
 more centrally concentrated than the nucleated dwarfs. 
They suggested that this result may reflect 
a sequential evolution from dwarfs to UCDs.
\citet{zha15} found that the UCDs near the M87 region show a stronger rotation and more radially biased orbits at large radii than M87 GCs, suggesting that UCDs have a dwarf galaxy origin.


	\begin{figure}
\epsscale{1}
\includegraphics[trim=10 20 10 20,clip,width=\columnwidth]{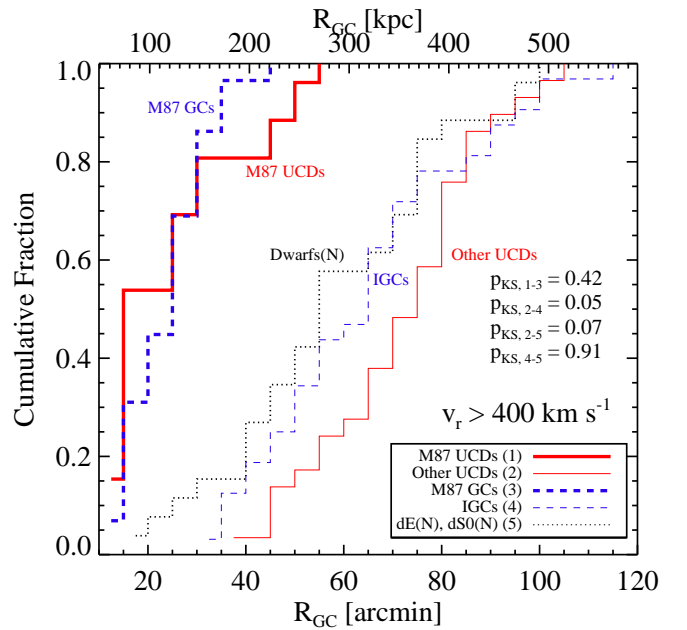}
\caption{Cumulative radial distribution of UCDs in the M87 region (thick solid line), the UCDs not in the M87 region (solid line), M87 GCs (thick dashed line), IGCs with \vrad $>$ 400 \kms (dashed line), and nucleated dwarf galaxies with \vrad $>$ 400 \kms in the survey region (dotted line, Kim et al. 2014).
\label{fig:cumnd}}
	\end{figure}
	\begin{figure*}
\epsscale{1}
\plotone{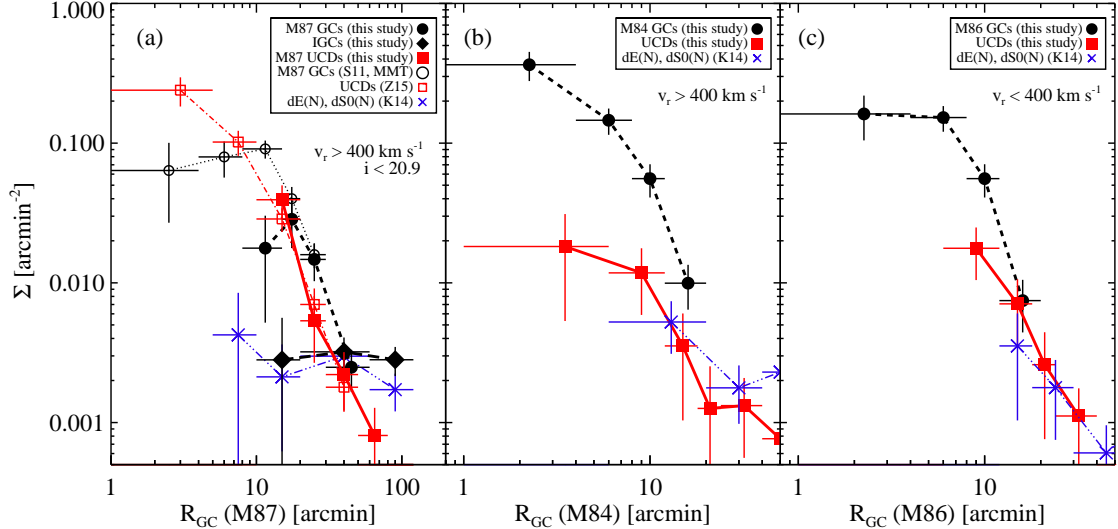}
\caption{(a) Radial number density profiles of M87 GCs/IGCs with $i < 20.9$ (circles), UCDs with $i < 20.9$ (squares), and nucleated dwarfs (crosses).
All objects have the radial velocities of \vrad $>$ 400 \kmsend.
The open and filled circles represent the M87 GCs observed with MMT/Hectospec in \citep{str11} and M87 GCs/IGCs in this study, respectively.
The open an filled squares represent the UCDs in \citep{zha15} and the UCDs in the M87 region confirmed in this study, respectively.
(b) Radial number density profile of M84 GCs (filled circles), UCDs with \vrad $>$ 400 \kms (filled squares), and nucleated dwarfs with \vrad $>$ 400 \kms (crosses) as a function of galactocentric distance from M84.
(c) Radial number density profile of M86 GCs (filled circles), UCDs with \vrad $<$ 400 \kms (filled squares), and nucleated dwarfs with \vrad $<$ 400 \kms (crosses) as a function of galactocentric distance from M86.
\label{fig:rd_comp}}
	\end{figure*}
We investigate the cumulative radial distribution of the UCDs, compared with those of M87 GCs, IGCs, and Virgo nucleated dwarfs (see Figure \ref{fig:cumnd}).  
We select nucleated dwarfs in the EVCC, and the dwarf galaxy population in the region covered by this study is dominated by nucleated dwarfs.
All these objects have radial velocities with \vrad $>$ 400 \kmsend, which belong to the main Virgo cluster.
We divide the UCDs into two groups (a group of UCDs in the M87 region (M87 UCDs, see \S3.1.2) and a group of the rest UCDs (Other UCDs). 
We estimate $p$-values from the two-sided Kolmogorov-Smirnov (K-S) test to compare their cumulative radial distributions.
The $p$-values indicate the probability that the two data come from the same parent population.
The $p$ value for the M87 UCDs and  M87 GCs  is 0.42. 
This indicates that the M87 GCs and M87 UCDs show a marginally different cumulative distribution. We could not make a concrete conclusion, because our survey region does not cover the innermost region of M87.
The $p$ values are derived to be 
 0.05 for the other UCDs and IGCs, and 0.07 for the other UCDs and nucleated dwarfs.
It means that the UCDs not in the M87 region have a clearly different radial distribution from that of IGCs and nucleated dwarfs.
Note that the radial distribution of the IGCs is similar to that of nucleated dwarfs with a confidence level of 91\%.


{\bf Figure \ref{fig:rd_comp}(a)} shows the radial number density profiles of the M87 GCs, IGCs, UCDs in the M87 region, and nucleated dwarfs. We included also the samples of GCs and UCDs spectrocopically confirmed in the previous studies \citep{str11, zha15}.
For the fair comparison, we select the GC and UCD samples with $i < 20.9$ and \vrad $>$ 400 \kms in the previous studies.
\citet{str11} compiled the GC sample with four different observations, but we select the GCs observed with MMT/Hectospec because of their wide spatial coverage.
Their MMT/Hectospec observations did not cover the innermost region of M87.
The radial number density profiles of the M87 GCs and UCDs confirmed in this study are well consistent in the range of 10$\arcmin <$ \rgc $<$ 40$\arcmin$ with those of M87 GCs in \citet{str11} and \citet{zha15}, respectively.
The GCs and UCDs show a clear central concentration on M87, while the IGCs and nucleated dwarfs do not.
The difference of the radial number density profiles between M87 GCs and UCDs is not significant.
{\bf Figure \ref{fig:rd_comp}(b)} shows the radial number density profiles of GCs, UCDs, and dwarfs with \vrad $>$ 400 \kms as a function of galactocentric distance from M84.
The GCs in M84 and UCDs show a strong central concentration on M84, but the nucleated dwarfs do not.
We notice that the excess on the cumulative radial number distribution of the UCDs at 80$\arcmin <$ \rgc $<$ 100$\arcmin$ is caused by the UCDs in M84 region.
The radial velocity dispersion of the UCDs with \vrad $>$ 400 \kms is 361 $\pm$ 24 \kmsend, which is much smaller than that of dwarf galaxies with \vrad $>$ 400 \kms ($\sigma = 608 \pm 84$ \kmsend).
The UCDs and IGCs with \vrad $>$ 400 \kms are more relaxed than the dwarfs as mentioned in \S4.3.

All UCDs with \vrad $<$ 400 \kms are located in the north-western region of M87 that is occupied by the M86 group (see Figure \ref{fig:spatial_v}(d)).
Figure \ref{fig:rd_comp}(c) shows the radial number density profiles of the M86 GCs, UCDs, and nucleated dwarfs with \vrad $<$ 400 \kms as a function of galactocentric distance from M86.
All these objects show a strong central concentration on M86, which indicates that the infalling structure is centered on M86.
The radial velocity dispersion of the UCDs with \vrad $<$ 400 \kms is 217 $\pm$ 41 \kmsend, which is similar to that of the dwarf galaxies with \vrad $<$ 400 \kms ($\sigma = 282 \pm 57$ \kmsend) and IGCs with \vrad $<$ 400 \kms ($\sigma = 172 \pm 33$ \kmsend).

In summary, we found two characteristics of the UCDs in the Virgo: 1) the UCDs show a stronger central concentration on massive galaxies (M87, M86, and M84) than the nucleated dwarfs, and 2) the radial velocity dispersion of the UCDs is similar to that of the GCs, but is much lower than that of the dwarf galaxies.
These two aspects of the UCDs are similar to those of the GC system, which implies that the UCDs in the Virgo may be simply the  bright-end GCs. However, recent studies on the GCs and UCDs in Virgo \citep{liu15,zha15} provide strong evidence supporting the dwarf galaxy origin of the UCDs: the difference in the cumulative radial distribution of the UCDs with and without diffuse envelope, and the difference in the kinematics of the UCDs and GCs around M87. Therefore more studies are needed to clarify the origin of the UCDs. 
Our UCD sample does not cover the entire azimuthal range of M87 and the numbers of the UCDs in M84 and M86 in this study are small. It is needed to use a larger number of UCDs to understand the origin of UCDs.

\section{SUMMARY}

We present the results from an extensive spectroscopic survey of GCs in the central region of the Virgo including the outermost region of M87.
The GCs, UCDs, foreground stars, and background galaxies are classified using several criteria including their radial velocities, sizes, and spectral morphology.
Finally, we identified 201 GCs and 55 UCDs from 910 spectroscopic targets.
The primary results of this paper are summarized as follows.

\begin{itemize}

\item
We detect 46 IGCs that do not belong to any individual bright galaxies in the Virgo, but wander in the intracluster region.
The total number of IGCs is expected to be about 2114 in the survey region.
The number density of the IGCs confirmed in this study is about 0.19 $\pm$ 0.02 arcmin$^{-2}$.
This is comparable with the number density expected from photometric surveys in the previous studies.

\item
The color distribution of IGCs shows two populations, blue (metal-poor) GCs and red (metal-rich) GCs, as in the case of M87 GCs. The blue IGCs mainly originate from dwarf galaxies, and some from massive galaxies. On the other hand, the red IGCs are only from intermediate-mass to high mass galaxies.

\item
The radial velocity distribution of IGCs does not follow a single Gaussian distribution.
There are two groups with mean radial velocities  \vrad = 1023 \kms and 36 \kmsend, respectively. 
The IGCs with high radial velocities are located in the main body of the Virgo, while the IGCs with low radial velocities may be infalling or outgoing populations.

\item
The radial velocity dispersion of the IGCs and UCDs with \vrad $>$ 400 \kms ($\sigma = 314 \pm 39$ \kms and 361 $\pm$ 24 \kmsend) is much lower than that of dwarfs with \vrad $>$ 400 \kms ($\sigma = 608 \pm 84$ \kmsend).
This indicates that the Virgo based on the kinematics of the IGCs and UCDs may be dynamically more relaxed systems than indicated by the dwarfs.


\item
Most of the UCDs confirmed in this study are located around the main galaxies in the central region of the Virgo, M84, M86, and M87, while the IGCs are not.

\end{itemize}

These results suggest that both IGCs and UCDs in the main body of the Virgo are dynamically cooler systems than dwarfs.
There are no IGCs with extremely high radial velocities (\vrad $>$ 2300 \kmsend) confirmed in this study unlike the dwarfs.
On the other hand, the IGCs and UCDs with low radial velocities are infalling toward or outgoing from the main body of the Virgo, but their spatial distributions are totally different.
The UCDs are thought to be involved in the infalling group centered on M86, but the origin of the IGCs that lie scattered in the survey region is unclear.
The size measurement of the compact IGCs will be very helpful to reduce the contamination due to foreground stars in the future.

\acknowledgments
We thank anonymous referee for her/his useful comments on the manuscript, and Daniel Fabricant for sharing his flux calibration routine of the MMT/Hectospec data.
This work was supported by the National Research Foundation of Korea (NRF) grant
funded by the Korea Government (MSIP) (No. 2012R1A4A1028713). 
This work was supported by K-GMT Science Program (PID: 14A-MMT003/2014A-UAO-G18) funded through Korean GMT Project operated by Korea Astronomy and Space Science Institute (KASI).
This paper uses data products produced by the OIR Telescope Data
   Center, supported by the Smithsonian Astrophysical Observatory.


\clearpage






\clearpage

\clearpage

\end{document}